\documentclass[pra,
               twocolumn,
               preprintnumbers,
               amsmath,
               amssymb,
               superscriptaddress,
               showpacs,
               longbibliographym,
               groupedaddress]{revtex4-2}

\usepackage{graphicx}
\usepackage{latexsym}
\usepackage{amsmath}
\usepackage{graphics}
\usepackage{amssymb}
\usepackage{layout}
\usepackage{verbatim}
\usepackage{amsfonts,epsfig,dsfont}
\usepackage{hyperref}
\hypersetup{
    colorlinks,
    citecolor=black,
    filecolor=black,
    linkcolor=black,
    urlcolor=black
}

\usepackage{nameref}
\usepackage{color}
\usepackage{ulem} 
\usepackage{soul}
\usepackage{cancel}
\usepackage{orcidlink}

\newcommand{\beq}{\begin{equation}}
\newcommand{\eeq}{\end  {equation}}

\usepackage{ulem} 

\makeatletter
\newcommand{\bal}{\@ifstar{\@bals}{\@bal}}
\def\@bals#1\eal{\begin{align*}#1\end{align*}}
\def\@bal#1\eal{\begin{align}#1\end{align}}
\makeatletter

\let\Re\undefined

\DeclareMathOperator{\Re}{\mathrm{Re}}

\begin{document} 
\title{
Limitations on the maximal level of entanglement
of two singlet-triplet qubits in GaAs quantum dots
}
\author{
              Igor Bragar\orcidlink{0000-0003-1321-2371}
       }
\affiliation{
              Institute of Physics,
              Polish Academy of Sciences,
              al.~Lotnik{\'o}w 32/46,
              PL~02-668 Warszawa, Poland
            }
\author{
              {\L}ukasz Cywi{\'n}ski\orcidlink{0000-0002-0162-7943}
       }
\email{
              lcyw@ifpan.edu.pl
      }
\affiliation{
              Institute of Physics,
              Polish Academy of Sciences,
              al.~Lotnik{\'o}w 32/46,
              PL~02-668 Warszawa, Poland
            }
\date{
              \today
     }
\begin{abstract}
  We analyze in detail a procedure of entangling of two singlet-triplet 
  ($S$-$T_{0}$) qubits
   operated in a regime 
  when energy associated with the magnetic field gradient, $\Delta B_{z}$,
  is an order of magnitude smaller than the exchange energy,  $J$,
  between singlet and triplet states
  [Shulman M. et al., Science \textbf{336}, 202 (2012)].
  We have studied theoretically a single $S$-$T_{0}$ qubit 
  in free induction decay and spin echo  experiments.
  We have obtained analytical expressions for time dependence of components of its Bloch vector 
  for quasistatical fluctuations of $\Delta B_{z}$ 
  and quasistatical or dynamical $1/f^{\beta}$-type fluctuations of $J$.
  We have then considered the impact of fluctuations of these parameters
  on the efficiency of the entangling procedure
  which uses an Ising-type coupling between two $S$-$T_{0}$ qubits.
  Particularly,
  we have obtained an analytical expression for evolution  of two qubits
  affected by $1/f^{\beta}$-type fluctuations of $J$.
  This expression indicates the maximal level of entanglement
  that can be generated by performing the entangling procedure.
  Our results deliver also an evidence
  that in the above-mentioned experiment
  $S$-$T_{0}$ qubits were affected by uncorrelated $1/f^{\beta}$ charge noises.
\end{abstract}
\maketitle

\hyphenation{
amp-li-tude
ap-pro-xi-ma-tion
ave-rage
bet-ween
cha-rac-te-ri-zing
co-he-rence
con-si-de-red
cont-rol
cons-tant
coup-led
coup-ling
con-si-de-ring
de-tu-ning
dy-na-mi-cal
dy-na-mi-cal-ly
ge-ne-rate
ge-ne-ra-tion
ener-gies
en-tang-led
en-tang-le-ment
en-tang-ling
exe-cute
ex-pe-ri-ment
ex-pe-ri-ments
fluc-tua-ting
fluc-tua-tions
ite-ra-tion
rea-lis-tic
rea-li-zed
re-so-nance
chal-len-ging
lo-ca-li-zed
ma-xi-mal
na-no-struc-tures
neg-lec-ting
nuc-lear
qua-si-sta-ti-cal-ly
rea-ching
rea-li-za-tion
re-pe-ti-tions
re-sul-ting
ope-ra-tor
pos-sible
pos-sib-ly
single
simp-lest
simp-ly
slowest
ty-pi-cal
vi-sible
}

\section{Introduction} 

Spin qubits based on gated quantum dots (QDs) 
\cite{Hanson_RMP07,Chatterjee_NRP21,Burkard_RevModPhys_2023}
can be initialized, coherently controlled, and read out.
Qubits based on a spin of a single electron 
\cite{Loss_PRA98}
localized in a QD can be controlled with electron-spin resonance techniques
\cite{Koppens_Nature06,Nowack_Science07,Pioro_NP08},
while two-qubit gates can be performed with the help of exchange interaction
\cite{Loss_PRA98,Burkard_PRB99,Li_PRB10},
which is controlled electrically
\cite{Petta_Science05,Brunner_PRL11,Veldhorst_Nature15,Reed_PRL16,Martins_PRL16,Watson_Nature18,Huang_Nature19}.
Implementation of single-spin control with ac electric
\cite{Nowack_Science07,Pioro_NP08,Laird_PRL07}
and magnetic 
\cite{Koppens_Nature06,Brunner_PRL11}
fields
is nevertheless experimentally challenging, especially in GaAs based QDs, in which interaction with nuclei 
\cite{Coish_pssb09,Cywinski_APPA11,Chekhovich_NM13,Yang_RPP17}
leads to significant broadening of electron spin resonance lines (this is in contrast to experimental situation in Si-based single-electron QDs 
\cite{Kawakami_PNAS16,Veldhorst_NN14,Yoneda_NN18}, for which nuclear noise can be removed by isotopic purification 
\cite{Yoneda_NN18,Struck_NPJQI20}).
Research on all-electrical control (possibly without ac fields)
is an active field. 
Such a control is possible for spin qubits based on a few
electrons localized in multiple quantum dots 
\cite{Chatterjee_NRP21}.
In a double quantum dot (DQD) containing two electrons
one can easily initialize and partially control a qubit,
whose logical states correspond to singlet $S$ and unpolarized triplet $T_{0}$
formed by the two electrons 
\cite{Petta_Science05}
(note that one can as well use two QDs with larger odd number of electrons
per dot 
\cite{Higginbotham_PRL14}),
and all-electrical ac control is also possible in a variety of other designs based on two
\cite{Kim_Nature14},
three
\cite{Medford_NN13,Medford_PRL13,Russ_PRL18}
and four
\cite{Sala_PRB17} electrons.
We focus here on the DQD-based two-electron singlet-triplet ($S$-$T_{0}$) qubit, for which full electrical control over the state of the qubit is possible
when the spin splittings of electrons localized in the two dots are distinct.
This can be achieved with creation of a gradient of nuclear spin polarization
\cite{Bluhm_PRL10,Foletti_NP09}
in nuclear spin rich material such as GaAs
or with the help of micromagnets creating a gradient of magnetic field
\cite{Brunner_PRL11}.

Creation of entanglement 
\cite{Horodecki_RMP09,Aolita_RPP15}
of two spin-based qubits is the next necessary step in development
of a QD spin qubit platform for quantum information processing.
For two single-spin qubits, exchange interaction leads
to creation of entangled two-qubit states 
\cite{Brunner_PRL11,Nowack_Science11,Watson_Nature18,Veldhorst_Nature15},
while in the case of $S$-$T_{0}$ qubits it is the electric dipole-dipole (capacitive) interaction
\cite{Taylor_NP05} 
that has been most commonly used for interqubit coupling
\cite{vanWeperen_PRL11,Shulman_Science12} (although exchange coupling could be used too 
\cite{Li_PRB12,Wardrop_PRB14,Cerfontaine_PRB20}).
Our focus in the present paper is on creation of entanglement
and its evolution due to capacitive interqubit interaction, 
as it was demonstrated experimentally for two $S$-$T_{0}$ qubits in Ref.~\cite{Shulman_Science12}. 

Interaction of qubits with their environments
that fluctuate in an uncontrolled manner leads to decoherence
\cite{Hornberger_LNP09} of their quantum states.
In this process, the entanglement,
which requires existence of a coherent superposition
of at least two product states of the two qubits,
is also destroyed 
\cite{Aolita_RPP15}.
For single electron spins in QDs,
the dominant cause of entanglement decay is their hyperfine interaction
with the nuclear baths 
\cite{Mazurek_PRA14,Bragar_PRB15}.
Decoherence of $S$-$T_{0}$ qubits, on the other hand,
can be dominated by the nuclear bath at low singlet-triplet splitting
\cite{Bluhm_NP10},
but for large splittings it is mostly caused by charge noise~\cite{Coish_PRB05,Hu_PRL06,Dial_PRL13}, 
with the nuclear-induced decay 
\cite{Yang_PRB08,Hung_PRB13}
possibly playing a role when the fluctuations of $S$-$T_{0}$ splitting are suppressed
\cite{Weiss_PRL12}.
Influence of quasistatic charge and nuclear noises 
on the simplest protocol of generation of entanglement 
of two capacitively coupled $S$-$T_{0}$ qubits was considered in Ref.~\cite{Wu_PRB17}.
In this paper, we consider a more involved protocol from Ref.~\cite{Shulman_Science12}
under the influence of dynamical charge noise. 

In the experiment 
\cite{Shulman_Science12},
the two $S$-$T_{0}$ qubits were initialized in a separable state,
and subsequently they were evolving
in the presence of a finite singlet-triplet splitting $J$.
With both qubits having nonzero $J$,
dipolar interaction between them leads to the appearance of an Ising-type interaction,
which in the decoherence-free case would lead to periodic creation
of maximal two-qubit entanglement.
In the experiment, only one period of such entanglement oscillation
(with entanglement reaching only a fraction of its maximal possible value)
is visible.
Furthermore,
for the entanglement to be nonzero
the two qubits have to be subjected to a spin echo refocusing pulse
that removes the influence of the slowest environmental fluctuations
on their evolution. 
The strong decoherence is significantly affecting the entanglement generation
and evolution. 
The goal of this paper is to understand
processes originated from nuclear polarization fluctuations and charge noise
affecting the singlet-triplet splittings 
which limit the ability to generate entangled two-qubit states
for experimental protocol performed in Ref.~\cite{Shulman_Science12}.
The main conclusion of our work is
that the spin echo protocol removes most of the influence
of the nuclear polarization fluctuations,
and the observed decoherence is caused by charge noise.
The experimental data from Ref.~\cite{Shulman_Science12}
are consistent with the assumption
that each qubit is subjected to $1/f^\beta$ charge noise
with $\beta\! \lesssim \! 1$
(as seen for a single $S$-$T_{0}$ qubit 
\cite{Dial_PRL13}),
and the noises affecting the two qubits are independent.
We also discuss the qualitative difference in the character
of the dominant decoherence process between the cases
of small $\beta \! \lesssim \! 1$
and
large $\beta \! > \! 1$
-- in the former case the imperfectly echoed-out single-qubit noise
has dominant influence,
while in the latter case of very low frequency noise the fluctuations
of two-qubit coupling are the main reason for imperfect entanglement.

The paper is organized as follows.
In Sec.~\ref{sec:1STq} we give  an overview of the physics of a single $S$-$T_{0}$ qubit,
and we discuss the influence of fluctuating external magnetic
or electric fields on the decoherence of such qubit seen
in the free induction decay (FID) as well as spin echo (SE) experiments.
In Sec.~\ref{sec:2STq}
we recall a procedure that has been designed for entangling two $S$-$T_{0}$ qubits~[\onlinecite{Shulman_Science12}]. 
We briefly discuss ways of entanglement quantification 
that have been applied to the system of two qubits 
in Sec.~\ref{sec:EntQuantif}.
In Sec.~\ref{sec:EntProcedureDecoh}
we then analyze the influence of the above-mentioned factors that lead to decoherence
on the efficiency of the procedure of entangling of two $S$-$T_{0}$ qubits.
We show here that a dynamically fluctuating electric field
affecting qubits' exchange splittings
may limit the maximal level of two-qubit state entanglement
by destroying two-qubit entangling gate or simply by dephasing of single qubits.
Finally, Sec.~\ref{sec:Conclusions} closes the paper with a discussion of conclusions on the nature of charge noise in the system studied in Ref.~\cite{Shulman_Science12} that one can draw by comparing the results of our calculations to the observations described there.
In the appendices we present
explicit expressions that describe the averaged values of $S$-$T_{0}$ qubit components
as functions of a duration of free induction decay and spin echo experiments
as well as attenuation factors
originated from $1/f^{\beta}$ dynamical noise of exchange splittings of the qubits.

\section{The physics of a Singlet-Triplet qubit} 
\label{sec:1STq}                                 

\subsection{The Hamiltonian and control over the qubit} 
In a singlet-triplet qubit, the quantum state is stored
in the joint spin state of two electrons in a DQD,
with one electron localized in each of the two dots,
the left (L) and the right (R) one
\cite{Taylor_NP05,Petta_Science05,Foletti_NP09,Dial_PRL13}.
The logical states of the qubit are
the singlet
$|S\rangle ~~=~
 \frac{1}{\sqrt{2}}(  \left| \uparrow_L   \downarrow_R \right\rangle
                    - \left| \downarrow_L \uparrow_R   \right\rangle)$
and
the spin unpolarized triplet
$|T_0\rangle ~=~
 \frac{1}{\sqrt{2}}(   \left| \uparrow_L   \downarrow_R \right\rangle
                     + \left| \downarrow_L \uparrow_R   \right\rangle)$.
The remaining two states,
spin polarized triplets,
$\left| \uparrow_L   \uparrow_R   \right\rangle$
and
$\left| \downarrow_L \downarrow_R \right\rangle$,
are split off by the constant magnetic field
applied in the plane of the nanostructure.
In the paper, we adopt the convention
that the Bloch sphere of $S$-$T_0$ qubit is defined in such a way
that state $|S\rangle$ ($|T_0\rangle$)
coincides with north (south) pole of the sphere
and the axis connecting these two points is the $z$ axis.

It was demonstrated in Ref.~\cite{Petta_Science05}
that it is possible to reliably initialize the two electron spins
in singlet state $|S\rangle$,
perform rotations around $z$ axis of the Bloch sphere,
as well as a read-out in a form of projective measurements onto $|S\rangle$.
All these operations can be realized
utilizing the fast control of the exchange interaction
that is achieved by applying proper voltage pulses to the metallic gates
on the surface of the sample.
The states $|S\rangle$ and $|T_0\rangle$ are
naturally split by energy $J$ due to the exchange interaction between electrons
in the DQD 
\cite{Burkard_PRB99,Li_PRB10}.
This energy difference can be influenced by controlling either
the energy difference between the ground single-electron states of the two dots
(the so-called detuning $\epsilon$) 
\cite{Petta_Science05,Taylor_PRB07},
or the height of the inter-dot barrier 
\cite{Loss_PRA98,Martins_PRL16}.
According to Ref.~\cite{Shulman_Science12},
the value of $J$ can be varied from much less than $1~\mu$eV to a few $\mu$eV
on a time scale of a nanosecond. 
We have then the following 
time-dependent and externally controlled term in  the Hamiltonian of the qubit
(in units of $\hbar$):
\begin{equation}
 \hat{H}_{J}(t) = J(t) \frac{\hat{\sigma}_{z}}{2} \,\, ,
\end{equation}
where $\hat{\sigma}_{z} = |S\rangle \langle S| - |T_0 \rangle \langle T_0 |$.

Rotations about the $x$ axis of the Bloch sphere
(or equivalently, the rotations
between $\left|S\right\rangle$ and $\left|T_0\right\rangle$ states)
present a greater challenge.
In Ref.~\cite{Foletti_NP09} it was realized with the help of an interdot gradient of electron spin splitting $\Delta B_{z}$ caused by difference of nuclear polarizations in the two dots.
If the local fields (either nuclear Overhauser fields, or magnetic fields from external magnets) in both dots were identical,
$|S\rangle$ and $|T_0\rangle$ states would not experience any dynamics,
since the phase acquired by spin-down state $\left|\downarrow_L\right\rangle$
(spin-up state $\left|\uparrow_L\right\rangle$)
of the electron in the $L$ dot would be cancelled by the spin-up state
$\left|\uparrow_R\right\rangle$
(spin down state $\left|\downarrow_R\right\rangle$)
of the electron in the second dot.
The field gradient breaks this symmetry,
which can be seen in the following example,
where we allow $|S\rangle$ to evolve for time~$t$
\begin{align}
 |S\rangle &= \frac{ 1 }{ \sqrt 2 }
             \left(   \left| \uparrow_L   \downarrow_R \right\rangle
                    - \left| \downarrow_L \uparrow_R   \right\rangle
             \right) \nonumber \\
           &\xrightarrow{t}
             \frac{ 1 }{ \sqrt 2 }
             \left(   \left| \uparrow_L   \downarrow_R \right\rangle
                    - \mathrm{e}^{i t {\Delta B_{z}}}
                      \left| \downarrow_L \uparrow_R   \right\rangle
             \right) \, .
 \label{eq:fromStoT}
\end{align}
We can see
that the initial state transforms back-and-forth
between singlet and the unpolarized triplet,
as the phase factor oscillates between $1$ and $-1$ with frequency set by $\Delta B_{z}$.
The field gradient $\Delta B_{z}$ contributes 
the following term
to the qubit's Hamiltonian:
\begin{equation}
\hat{H}_{\Delta B_{z}} = \Delta B_{z} \frac{\hat{\sigma}_{x}}{2} \,\, ,
\end{equation}
where $\hat\sigma_x = \left| S   \right\rangle \left\langle T_0 \right|
                    + \left| T_0 \right\rangle \left\langle S   \right|$.

The difficulty in the realization of precise singlet-to-triplet transitions lies
in the fact that the necessary field gradient, in most cases,
is generated by the slowly fluctuating Overhauser field
established by the nuclear spins of the atoms comprising the sample
\cite{Reilly_PRL08}.
Therefore, $\Delta B_{z}$ must be treated as given
(in fact, it must be measured beforehand, 
and the value of $\Delta B_{z}$ varies in different repetitions of the experiment),
hence, it would induce an ongoing transition
between $\left|S\right\rangle$ and $\left|T_0\right\rangle$,
which is undesirable in the context of the entangling procedure.
Instead, such a procedure requires precise state transformations on demand.
For example,
if one desires to execute a rotation
from $\left| S   \right\rangle$
to   $\left| T_0 \right\rangle$
at a given moment $t^{\prime}$ it would be ideal
if $\Delta B_{z}$ could be turned on only for an interval
$[t^{\prime} , t^{\prime} + \pi/\Delta B_{z} ]$,
so that the phase in Eq.~(\ref{eq:fromStoT}) is
$\mathrm{e}^{i \Delta B_{z} \cdot ( t^{\prime}+\frac{ \pi}{\Delta B_{z}} - t^{\prime})} =
\mathrm{e}^{i\pi}=-1$.
Since there is no practical way to change the value of $\Delta B_{z}$
during a single realization of the procedure,
one cannot simply turn off the gradient and stop the transition.
Nevertheless,
the transition can be effectively blocked by overshadowing $\Delta B_{z}$
with strong enough splitting $J$,
which can be controlled with relative ease, and it can be switched on and off at will.
In order to realize the idealized scenario such as the one described above,
the time scale on which $J$ is manipulated
must be the shortest time scale in the problem.
In particular,
this time scale must be much shorter
than the period of rotations around $x$-axis $1/\Delta B_{z}$
as well as the period of $z$-rotations
set by the maximal value of singlet-triplet splitting 
$1/J_\mathrm{max}$.
Note that pulse-shaping techniques for gate error mitigation were derived for $S$-$T_{0}$ qubits 
\cite{Wang_NatComm12,Kestner_PRL13,Wang_PRA14}, but their implementation has proven to be challenging so far, and we focus here on the simplest control scheme used in experiment on creation of entanglement in Ref.~\cite{Shulman_Science12}. Let us also note that entanglement of two $S$-$T_{0}$ qubits in the situation in which $\Delta B_z$ is always larger than $J$ was generated  using a scheme involving ac control of $J$ 
\cite{Nichol_npjqi17},
which is distinct from the one used in Ref.~\cite{Shulman_Science12} and analyzed here.

\subsection{Decoherence of a single $S$-$T_{0}$ qubit} 
\label{sec:1STq_decoherence}                           

\subsubsection{The nature of noisy terms in the Hamiltonian}
A qubit evolving with nominally constant $\Delta B_{z}$ and $J$ is undergoing decoherence
due to uncontrolled fluctuations of these parameters.
When the magnetic field gradient $\Delta B_{z}$ is due to the difference of the $z$ components
of nuclear Overhauser fields in the two dots
(which is the case on which we focus here),
the main mechanism responsible for its fluctuations is
the spin diffusion process~\cite{Reilly_PRL08,Reilly_PRL10}
caused by dipolar interactions between the nuclear spins.
Note that we focus here on $J\! \gg \Delta B_{z}$ regime,
in which only the large-amplitude classical fluctuations
of the Overhauser fields can affect the qubit.
This is in contrast to $J\! \ll \! \Delta B_{z}$ case,
in which a quantum treatment of nuclear fluctuations is necessary
\cite{Yao_PRB06,Witzel_PRB06,Witzel_PRB08}.
The large-amplitude fluctuations have a characteristic decorrelation time scale
of about one second 
\cite{Reilly_PRL08,Reilly_PRL10},
which is much longer than the time scale of a single run of the experiment,
i.e.~a single repetition
of qubit's initialization -- evolution -- measurement cycle.
Therefore, $\Delta B_{z}(t)$ can be treated as quasistatic
\cite{Taylor_QIP06,Cywinski_APPA11,Hung_PRB13},
i.e.~it is considered as a constant (i.e.~time independent) random variable
with certain probability distribution $p(\Delta B_{z})$.
For a large number of nuclei, this distribution is normal~\cite{Merkulov_PRB02},
with the average value
$\overline{\Delta B_z}$
and dispersion (standard deviation)
$\sigma_{\Delta B_{z}} = \sqrt{\overline{\Delta B_z^2}-(\overline{\Delta B_z})^2}$.
Within this approximation,
the result of the experiments are interpreted as follows.
Given the initial density operator of the qubit,
$\hat\rho(0) = \hat \rho_\mathrm{ini}$,
its evolution during the experiment run number $n$,
is described by the qubit Hamiltonian 
$\hat H_{\mathrm{q}}(t) = \hat{H}_{J}(t) + \hat{H}_{\Delta B_{z}}$ 
with an unknown, but constant, value of $\Delta B_{z} = \Delta B_{z}^{n}$
drawn from the distribution $p(\Delta B_{z})$,
\begin{align}
 \hat \rho(t|\Delta B_{z}^n) 
 &=
 \mathrm{e}^{-i t(J\hat{\sigma}_{z} + \Delta B_{z} \hat{\sigma}_{x})/2}
 \hat \rho_\mathrm{ini}
 \nonumber \\
 &\times
 \mathrm{e}^{ i t(J\hat{\sigma}_{z} + \Delta B_{z} \hat{\sigma}_{x})/2}
 \Big|_{\Delta B_{z}=\Delta B_{z}^n} \, .
\end{align}
The final results are derived from the whole series
of $N$ experiment repetitions with the same initial state.
Therefore, the expectation values of the measured quantities are calculated
with the help of the averaged density operator according to
\begin{align}
 \hat \rho_\mathrm{fin}(t)
 &=
 \frac{1}{N}\sum_{n=1}^N \hat\rho(t|\Delta B_{z}^n)
 \nonumber \\
 &
 \xrightarrow{N\to\infty} \int_{-\infty}^{\infty} d(\Delta B_{z}) p(\Delta B_{z}) \hat\rho(t|\Delta B_{z})
  \nonumber \\
 &
 = \overline{\hat\rho(t| \Delta B_{z})} \, .
\end{align}

On the other hand,
the exchange splitting $J(t)$ changes due to fluctuations
of local electric fields,
i.e.~due to charge noise, which is ubiquitous in semiconductor nanostructures.
The charge noise typically has its spectral weight
concentrated at low frequencies.
Contributions of noise at frequencies corresponding to the inverse of typical qubit coherence are typically negligible when considering free evolution of the qubit (hence the noise can be treated then as quasistatic), but they have to be taken into account when modeling the spin echo experiment~\cite{Dial_PRL13},
in which the influence of the lowest-frequency noise is removed, and coherence times are longer.
In order to model the experimental data, one has to average the qubit's evolution 
over many realizations of the stochastic process $J(t)$.
If the noise statistics is assumed to be Gaussian
(which is natural for noise consisting of many independent contributions),
and if the evolution can be treated in the pure dephasing approximation 
(i.e.~neglecting $\Delta B_{z} \hat{\sigma}_{x}$ term when $\Delta B_{z} \! \ll \! J(t)$),
the averaging can be done analytically.
In the case of non-Gaussian noise and when keeping the general form
of the Hamiltonian, one has to resort to numerical simulations~\cite{Ramon_PRB12,Ramon_PRB15}.

It was shown
\cite{Dial_PRL13}
that the GaAs/AlGaAs $S$-$T_{0}$ qubit is affected by noise
having power spectral density of $1/f^{\beta}$ form
with $\beta \! \approx \! 0.7$ in the range of frequencies relevant for correct description of spin echo signal.
It is unclear if this value of $\beta$ is specific to this material
or only to the  device used in that experiment.

\subsubsection{Decoherence during free evolution of the qubit}
We now assume that the initial state of a qubit is
$|{-y}\rangle = \frac{1}{\sqrt{2}}(|S\rangle - i |T_0\rangle)$.
This choice is connected with the entangling procedure
used in Ref.~\cite{Shulman_Science12},
in which this single-qubit state is used as the initial one for both qubits.
In a free induction decay (FID) experiment, the qubit undergoes evolution without any manipulations between its initialization and the coherence readout at time $\tau$.
With fixed values of $J$ and $\Delta B_{z}$,
the expectation values of qubit observables
$\langle \hat \sigma_i^{\mathrm{FID}}(\tau) \rangle =
 \langle {-y} | \hat{\sigma}_i(\tau) | {-y} \rangle$
are given by
\begin{align}
 \langle \hat \sigma_x^{\mathrm{FID}}(\tau) \rangle & =
 \frac{J}{\sqrt{\Delta B_{z}^2 + J^2}} \sin \Big( \sqrt{\Delta B_{z}^2 + J^2} \tau \Big) \,\, , \\
 \langle \hat \sigma_y^{\mathrm{FID}}(\tau) \rangle & =
 -\cos\Big( \sqrt{\Delta B_{z}^2 +J^2} \tau \Big)  \,\, ,\\
 \langle \hat \sigma_z^{\mathrm{FID}}(\tau) \rangle & =
 -\frac{\Delta B_{z}}{\sqrt{\Delta B_{z}^2 +J^2}} \sin\Big( \sqrt{\Delta B_{z}^2 +J^2} \tau \Big) \,\, .
\end{align}
The qubit evolves under the influence of $J \! \gg  \! \Delta B_{z}$,
so to the lowest order in $\Delta B_{z}/J$ the initial amplitude of $x$ and $y$ signals
is $\approx 1$,
while the $z$ signal has amplitude $\approx \! \Delta B_{z}/J$.
In the presence of fluctuations of both $J$ and $\Delta B_{z}$
all these signals will average to zero at long times,
at which the arguments of the oscillatory functions
taken from an appropriate distribution will have relative phases
randomly distributed between $0$ and $2\pi$.

We can average the above expressions over $\Delta B_{z}$ and $J$ assuming Gaussian
and quasistatic fluctuations of either of these parameters.
We focus on the largest observable,
$\langle \hat \sigma_y^{\mathrm{FID}}(\tau) \rangle$,
and we approximate
$\sqrt{\Delta B_{z}^2 + J^2}\tau \approx J \tau + \Delta B_{z}^2 \tau/2J$,
which is valid for $J \gg \Delta B_{z}$ and  for short durations $\tau  \ll 8J^3/\Delta B_{z}^4$.
For $J$ noise we get then a simple Gaussian decay:
\begin{equation}
 \left \langle \langle \hat \sigma_y^{\mathrm{FID}}(\tau) \rangle \right \rangle_{J}  \approx 
 -\frac{1}{2} \exp \Big[- ( \tau/T^{*}_{2,J}) ^2  \Big]
 \cos(\bar J \tau) \,\, ,
 \label{eq:SingleQubit_y_J}
\end{equation}
where $\bar{J}$ is the average value of $J$ and the decay time scale is
\begin{equation}
 T^{*}_{2,J} = \frac{\sqrt{2}}{\sigma_{J}} \,\, ,
 \label{eq:T2starJ}
\end{equation}
where $\sigma_{J}$ is the standard deviation of $J$.

Averaging over $\Delta B_{z}$ gives
\begin{align}
 \left \langle \langle \hat \sigma_y^{\mathrm{FID}}(\tau) \rangle \right \rangle_{\Delta B_{z}} & \approx 
 - \exp\left( -\frac{\tau^2 \overline {\Delta B_{z}}^2 \sigma_{\Delta B_{z}}^2}{2 \left( J^2 + (\sigma_{\Delta B_{z}}^2 \tau )^2\right)} \right) \nonumber \\ 
&\times \frac{1}{\big( 1+(\sigma_{\Delta B_{z}}^2 \tau/J)^2 \big) ^{1/4}}
 \nonumber \\
&\times
\cos \Big( r(\tau)+s(\tau) \Big), \label{eq:FIDB}
\end{align}
where 
$
r(\tau) = \frac{1}{2} \arctan \left(\sigma_{\Delta B_{z}}^2 \tau/J\right)
$,
and   
$
s(\tau) = J\tau \Big( \overline{\Delta B_{z}}^2 +2 J^2 + 2 ( \sigma_{\Delta B_{z}}^2 \tau )^2 \Big)
/
\Big(2 \big( J^2+( \sigma_{\Delta B_{z}}^2 \tau )^2 \big) \Big)
$.
The decay envelope is then a product of two factors.
The first one dominates the decay when
$\overline{\Delta B_{z}} \! \gg \! \sigma_{\Delta B_{z}}$, 
i.e.~in the situation in which a finite $\Delta B_{z}$ is used for coherent control of the qubit. 
Then at long $\tau$ this factor saturates at 
$\exp\big( -\overline{\Delta B_{z}}^2/2\sigma^{2}_{\Delta B_{z}} \big) \! \ll \! 1$, 
and the qubit loses most of its coherence at time scale $\tau \! \ll \! J/\sigma^2_{\Delta B_{z}}$, 
at which the factor can be approximated as $\exp \Big( -(\tau/T_{2,\Delta B_{z}}^{*})^2 \Big)$,
where
\begin{equation}
 T^{*}_{2,\Delta B_{z}} \! = \! \frac{\sqrt{2}}{\sigma_{\Delta B_{z}}}
                          \cdot \frac{J}       {\overline{\Delta B_{z}}} \,\, .
 \label{eq:T2starDB}
\end{equation}
On the other hand, when $\overline{\Delta B_{z}} \! \ll \! \sigma_{\Delta B_{z}}$,
the second factor dominates,
and the signal envelope decays in power law fashion $\propto \! \sqrt{\tau_{\Delta B_{z}}/ \tau}$,
where the characteristic time $\tau_{\Delta B_{z}} \! = \! J/\sigma^2_{\Delta B_{z}}$.

It is important to note now that in the regime of
$\bar{J} \! \gg \! \overline{\Delta B_{z}}$
and
$\overline{\Delta B_{z}} \! \gg \! \sigma_{\Delta B_{z}}$
(that is relevant for experiments of interest in this paper),
we typically have
$\sigma_{J} \! \gg \! \sigma_{\Delta B_{z}}$.
This is due to the observed relation between $J$ and detuning $\epsilon$:
$J \! \approx \! J_{0}\exp(\epsilon/\epsilon_{0})$,
and the fact that the noise in $J$ comes mostly from fluctuations of $\epsilon$.
We have thus $\delta J/J \sim \delta \epsilon/\epsilon_{0}$,
so for constant level of detuning noise the standard deviation $\sigma_{J}$ increases with $J$
\cite{Dial_PRL13}.
It is then a reasonable assumption to neglect the effect
of fluctuations of $\Delta B_{z}$ in the calculation of decoherence.
Furthermore,
for
$\bar{J} \! \gg \! \overline{\Delta B_{z}}$
the main effect of
$\overline{\Delta B_{z}}\hat{\sigma}_{x}$ term is a slight tilt in $xOz$ plane of the axis about which
the qubit's Bloch vector is precessing.
Neglecting this effect, we arrive at the pure dephasing approximation
to the qubit's Hamiltonian:
\begin{equation}
 \hat{H}(t) \approx \left( \bar{J} + \delta J(t) \right) \frac{\hat{\sigma}_{z}}{2} \,\, ,
 \label{eq:Hpd}
\end{equation}
where we have now included the time dependence of $J$ noise.
The off-diagonal element of the qubit's density operator is given by
\begin{align}
 \rho_{ST_{0}}(\tau) &=
 \rho_{ST_{0}}(0) \,
 \mathrm{e}^{ -i \bar{J} \tau }
 \nonumber \\ 
 &\times
 \left \langle
  \exp \left( -i \int_{0}^{\infty} f_{\mathrm{FID}}(t;\tau) \delta J(t) \mathrm{d}t \right)
  \right
 \rangle_{\delta J} \,\, ,
 \label{eq:rhoSTdef}
\end{align}
where $\langle ... \rangle_{\delta J}$ denotes averaging
over different realizations of $\delta J(t)$ noise,
and
$ f_{\mathrm{FID}}(t;\tau)\! = \! \theta(t)\theta(\tau-t)$
is the FID time-domain filter function, 
where $\theta(t)$ is Heaviside step function.
The transverse components of the qubit state are given by
$\left \langle \langle \hat \sigma_x^{\mathrm{FID}}(\tau) \rangle \right \rangle \! =\!
2\mathrm{Re}\rho_{ST_{0}}(\tau)$
and
$\left \langle \langle \hat \sigma_y^{\mathrm{FID}}(\tau) \rangle \right \rangle \! =\!
2\mathrm{Im}\rho_{ST_{0}}(\tau)$.
For Gaussian $\delta J(t)$ noise
only the second cumulant of the random phase is nonzero~\cite{deSousa_TAP09,Cywinski_PRB08,Szankowski_JPCM17},
and we have a closed formula for coherence:
\begin{equation}
 \rho_{ST_{0}}(\tau) =
 \rho_{ST_{0}}(0) \, \mathrm{e}^{-i\bar{J} \tau}
                  \, \mathrm{e}^{-\chi_{\mathrm{FID}}(\tau)} \,\, ,
 \label{eq:rhoSTchi}
\end{equation}
in which the attenuation factor $\chi_{\mathrm{FID}}$ 
is defined in the following way
\begin{align}
 \chi_{\mathrm{FID}} & =
 \int_{0}^{\infty} \mathrm{d}t_{1}
 \int_{0}^{\infty} \mathrm{d}t_{2}
 f_{\mathrm{FID}}(t_{1})
 f_{\mathrm{FID}}(t_{2})
 \left\langle \delta J(t_{1})\delta J(t_{2}) 
 \right\rangle_{\delta J} \nonumber \\ & =
 \int_{0}^{\infty} S(\omega)
                   |\tilde{f}_{\mathrm{FID}}(\omega)|^{2}
                   \frac{\mathrm{d}\omega}{\pi} \,\, , \label{eq:chiFID}
\end{align}
where in the second line we have assumed that the noise is stationary,
so that its autocorrelation function is
$C(t_{1}-t_{2}) \! = \!
\left\langle \delta J(t_{1}) \delta J(t_{2}) \right\rangle_{\delta J}$.
The spectral density of the noise is
$S(\omega) \! = \!
\int_{-\infty}^{\infty} C(t) \mathrm{e}^{i\omega t}\mathrm{d}t$
and the frequency domain filter function 
\cite{deSousa_TAP09,Cywinski_PRB08} is
\begin{align}
 \tilde{f}_{\mathrm{FID}}(\omega)
 &= \int_{-\infty}^{\infty} f_{\mathrm{FID}}(t;\tau) \mathrm{e}^{i\omega t} \mathrm{d}t
  \nonumber \\
 &
 =  2 \sin^2 \left( \frac{\omega \tau}{2} \right) \frac{1}{\omega^2} \,\, .
\end{align}
In the case of $S(\omega) \! \sim \! \frac{1}{f^\beta}$ with $\beta = 0.7$,
the attenuation factor defined in Eq.~(\ref{eq:chiFID}) depends on $\tau$ as a power function:
$
\chi_{\mathrm{FID}} \propto \tau^{1.7}
$
(see Appendix \ref{app:chi} for a detailed calculation).

Note that for $\beta \! \geqslant \! 1$ the integral in Eq.~(\ref{eq:chiFID}) diverges.
However, in a real experimental setting, the total time of data acquisition, $T_M$,
involving many repetitions of cycles of qubit initialization, evolution for time $\tau$,
and measurement, sets the low-frequency cutoff,
$\omega_{\mathrm{min}} \! \sim \! 1/T_M$,
for frequencies of noise that actually affect the qubit~\cite{Szankowski_JPCM17,Dial_PRL13,Struck_NPJQI20}.
Consequently, the lower limit of the integral in Eq.~(\ref{eq:chiFID}) should be
$\omega_{\mathrm{min}} > 0$ instead of zero,
making the attenuation factor finite.
In this paper we are setting $\omega_{\mathrm{min}}/2\pi \! = \! 1$~mHz,
corresponding to $T_M$ in a perfectly realistic range of tens of minutes.

\begin{figure}[t]
\centering
  \includegraphics[width=\linewidth]{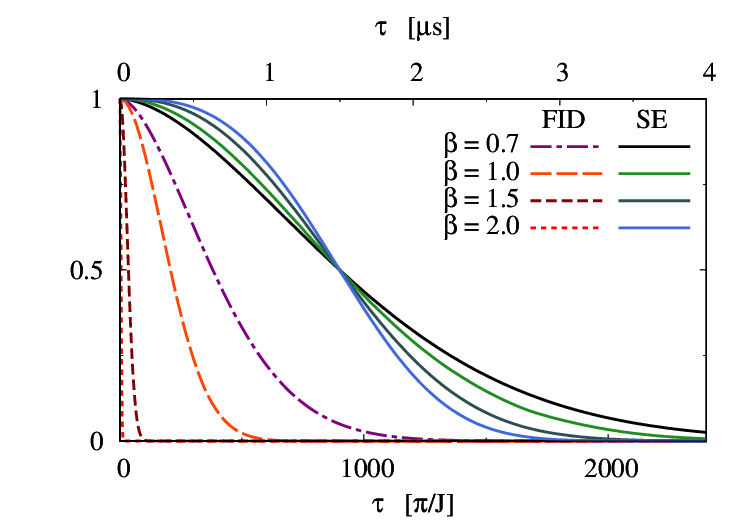}
   \caption{ (Color online)~FID and SE signals of a single $S$-$T_{0}$ qubit
             for the case of $1/f^{\beta}$ noise in exchange splitting $J$, with fluctuations in $\Delta B_{z}$ neglected.
             The power of the noise was chosen to be such 
             that ensures decay of SE signal as observed in the experiment
             \cite{Dial_PRL13} 
             (such a power that leads to half-decay of SE signal in 1.5 $\mu s$ for $J\! \approx \! 1$ $\mu$eV).
             Note that the improvement in coherence time due to echo relative to FID is larger
             when $\beta$ is larger,
             i.e.~when there is more noise power at the lowest frequencies.}
  \label{fig:FID_SE_dynamical}
\end{figure}

\subsubsection{Decoherence in spin echo protocol}
The influence of very low frequency noise on qubit dephasing can be removed
by employing a spin echo protocol 
\cite{deSousa_TAP09},
in which a qubit is subjected to a $\pi$ rotation
about an axis perpendicular to the axis along which the noise is coupled
(here we focus on $\pi$ rotations about the $x$ axis)
at time $t = \frac{\tau}{2}$, and the coherence is read out at the final time $t=\tau$.

Within the pure dephasing approximation introduced above,
and for perfect short ($\delta$ function like) pulses,
the calculation of echoed coherence signal 
as a function of duration $\tau$ of the echo procedure
amounts to replacing the $f_{\mathrm{FID}}(t)$ function
in Eq.~(\ref{eq:rhoSTdef}) by the $f_{\mathrm{SE}}(t)$ function
which is nonzero for $t\in[0,\tau]$ and changes its value from $1$ to $-1$
at $t=\tau/2$, see e.g.~Refs.~\cite{deSousa_TAP09,Szankowski_JPCM17}.
The coherence at the final readout time is given by a formula analogous
to Eq.~(\ref{eq:rhoSTchi}),
only with $\chi_{\mathrm{FID}}(\tau)$ replaced by $\chi_{\mathrm{SE}}(\tau)$
in which
$\tilde{f}_{\mathrm{SE}}(\omega) = 4\sin^{4}\frac{\omega \tau}{4} / 
\omega^2$ appears.
The latter filter function strongly suppresses very low $\omega$ contribution
(precisely from $\omega \! \ll \! 4\tau$ range)
to the attenuation factor $\chi_{\mathrm{SE}}(\tau)$.
For quasistatic charge noise we have
$S(\omega) \! \approx \! \sigma_{J}\delta(\omega)$,
and the echo protocol leads to a complete recovery of the initial coherence.

For $J$ noise with nontrivial spectrum,
but with a lot of noise power at low frequencies,
the echo decay time is expected to be much longer than the FID decay time,
see Fig.~\ref{fig:FID_SE_dynamical} for illustration.
Recall that our justification for the neglect of quasistatic $\Delta B_{z}$ fluctuations was 
that in the experimentally relevant parameter regime
they lead to much slower FID decay than the $J$ fluctuations.
Echo-induced suppression of $J$ noise effects could be suspected of leading
to breakdown of that assumption.
This is not the case:
the echo protocol is also strongly suppressing the effects
of quasistatic transverse noise,
provided that the effective transverse field $\Delta B_{z}$ is much smaller
than the typical longitudinal field.
For given values of $J$ and $\Delta B_{z}$ we have 
\begin{align}
  \langle \hat \sigma_y^{\mathrm{SE}}(\tau) \rangle
  &= \langle {-y} | \mathrm{e}^{ i \hat H \frac{\tau}{2}} ( i \hat \sigma_x)\mathrm{e}^{ i \hat H \frac{\tau}{2}}
  \hat \sigma_y
  \nonumber \\
  &\times
  \mathrm{e}^{-i \hat H \frac{\tau}{2}} (-i \hat \sigma_x)\mathrm{e}^{-i \hat H \frac{\tau}{2}}
  | {-y} \rangle \nonumber \\
  &=\frac{1}{\Delta B_{z}^2 + J^2}
  \Big( J^2 + \Delta B_{z}^2 \cos \Big( \sqrt{\Delta B_{z}^2 +J^2} \tau \Big) \Big) \,\, .
\end{align}
As before,
this expression can be analytically averaged over $\Delta B_{z}$
for short durations $\tau  \ll 8J^3/\Delta B_{z}^4$.
The full result is more complicated than Eq.~(\ref{eq:FIDB})
(see Appendix \ref{app:single} for full expression),
but for $\sigma^{2}_{\Delta B_{z}}\tau/J \! \gg \! (\Delta B_{z}/\sigma_{\Delta B_{z}})^2$
one can arrive at
\begin{align}
 \left \langle \langle \hat \sigma_y^{\mathrm{SE}}(\tau) \rangle \right \rangle_{\Delta B_{z}}
 &\approx
 \frac{J^2}{J^2+\overline{\Delta B_{z}}^2}
 \nonumber \\ 
 &
 + \frac{\sigma^2_{\Delta B_{z}}}{2 \left(J^2+\overline{\Delta B_{z}}^2\right)}
   \left(\frac{J}{\sigma^2_{\Delta B_{z}}\tau}\right )^{3/2}
   \nonumber \\
 &\times
 \cos \left( J \tau \right) \,\, ,
\end{align}
where we see that at long times the only effect of quasistatic transverse noise
is decrease of the coherence signal
by an amount $\approx (\overline{\Delta B_{z}} / J)^2\! \ll \! 1$ compared to its initial value.

The echo signal averaged over quasistatic fluctuations of $J$ looks similarly
\begin{align}
\langle \langle \hat \sigma_y^{\mathrm{SE}} (\tau) \rangle \rangle_{J}
&\approx
\frac{\bar J^2 + \sigma_J^2}{\bar J^2 + \Delta B_{z}^2}
 \nonumber \\
&
+\frac{\Delta B_{z}^2}{\bar J^2 + \Delta B_{z}^2} \exp\left(- \frac{\sigma_J^2 \tau^2}{2}\right)
 \nonumber \\
&\times
\cos \left( \bar J \tau \right).
\end{align}
For long times, 
the signal $\langle \langle \sigma_y^{\mathrm{SE}} (\tau) \rangle \rangle_{J}$
remains close to its initial value
${(\bar J^2 + \sigma_J^2)}/{(\bar J^2 + \Delta B_{z}^2)} \approx 1$.

\section{The procedure for entangling two $S$-$T_{0}$ qubits} 
\label{sec:2STq}                                      

Now we proceed with the description of the procedure designed
in Ref.~\cite{Shulman_Science12}
aimed to create maximally entangled two-qubit states
out of an initial product state.
Here, let us focus on an idealized setting in which both $J$ and $\Delta B_z$ are piecewise-constant and fully controlled, so that no averaging over their values is performed. Of course, in reality only $J$ is controlled, and it furthermore fluctuates  -- and the consequences of this will be the main subject of subsequent sections.

Entanglement generation requires some kind of qubit-qubit interaction, and in the case of $S$-$T_{0}$ qubits one utilizes the coupling between electric dipoles
induced by state-dependent charge distributions in each DQD 
\cite{Taylor_NP05}.
Only if both qubits are in state $\left|S\right\rangle$, and their exchange splittings are finite, the charge distributions
are asymmetric (due to mixing of the singlet state relevant here with a singlet state of two electrons localized in a single dot),
and hence, each DQD possesses a nonzero electric dipole moment.
Therefore,
the effective Hamiltonian of this interaction
is given by
(in units of $\hbar$)
\begin{equation}
 \hat H_{\mathrm{int}} =
 \frac{1}{4} J_{12} \left| S S \right\rangle \left\langle S S \right| =
 \frac{1}{4} J_{12} (\hat{\sigma}_z + \hat{\mathds{1}})
            \otimes (\hat{\sigma}_z + \hat{\mathds{1}}) \, .
\label{eq:H_int}
\end{equation}
It was established empirically in Ref.~\cite{Shulman_Science12}
that for values of splittings $J_i$ used there
the strength of the interaction is given by
\begin{equation}
    J_{12} =  \frac{J_1 J_2}{K} \,\, ,
\label{eq:J12}
\end{equation}
where parameter $K$ is a constant.
Thus,
the control over the exchange splittings (described in the previous section)
simultaneously allows to modify the value of the coupling $J_{12}$.
Note that the fact that $J_{12} \propto J_1 J_2$ exposes single-qubit gates to crosstalk
when nearby $S$-$T_{0}$ qubits have finite $J_i$ splittings,
and methods for dealing with this issue have been discussed
\cite{Buterakos_PRB18}.
Furthermore,
configuration interaction calculations have suggested the existence of parameter regions
for two capacitively coupled $S$-$T_{0}$ qubits
in which the relation between $J_{12}$ and $J_i$ is more complicated,
leading e.g.~to predictions of ``sweet spots''
at which charge-noise induced fluctuations of $J_{i}$ and/or $J_{12}$ are suppressed
\cite{Nielsen_PRB12,Chan_PRB21}.
Here we focus on noisy behavior of $J_i$ measured in Ref.~\cite{Dial_PRL13}
and the resulting noisy behavior of $J_{12}$ following from Eq.~(\ref{eq:J12}).

\begin{figure}[t]
\centering
 \includegraphics[width=\linewidth]{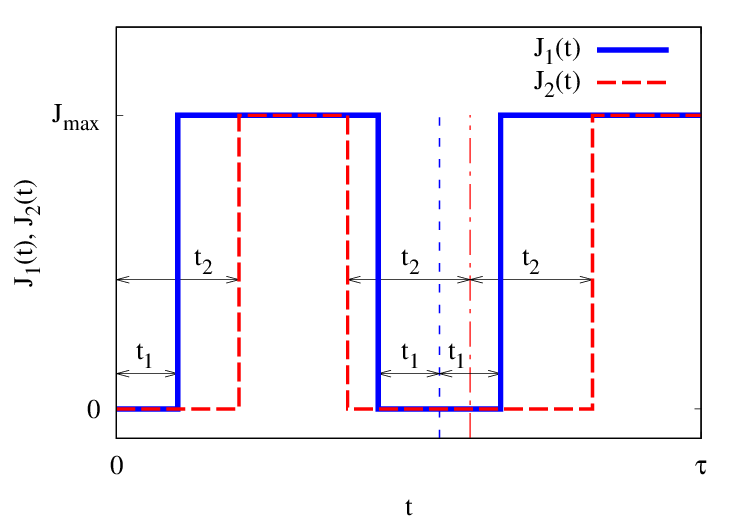}
 \caption{
           (Color online)~The temporal control of the exchange splittings $J_1(t), ~J_2(t)$
           during execution of the entangling procedure
           when the magnetic field gradients $\Delta B_{z, 1} \neq \Delta B_{z, 2}$.
           Thin vertical blue dashed and red dash-dotted lines
           are positioned at the middle of the durations
           of $\pi$ pulses
           for qubit 1 and 2, respectively.
         }
 \label{fig:procedure}
\end{figure}

The procedure consists of the following steps.
Before each run
the value of the gradients of magnetic field $\Delta B_{z,i}$
in each DQD is established
(e.g.~by a measurement or even setting the value with a dedicated procedure).
Then each $S$-$T_{0}$ qubit is independently initialized
in $\left|S\right\rangle$ state, yielding a separable two-qubit state
\begin{equation}
 \left| \psi(t=0) \right\rangle =      \left| S \right\rangle \otimes \left|S \right\rangle
                                \equiv \left| S S \right\rangle \,.
\end{equation}
For each qubit,
the exchange splitting is turned off for a time interval corresponding to $\frac{\pi}{2}$ rotation around $x$ axis
due to magnetic gradients.
Generally $\Delta B_{z,1} \neq \Delta B_{z,2}$
(for concreteness, suppose that $\Delta B_{z,1} \geqslant \Delta B_{z,2}$)
so that each splitting has to be kept turned off for different time interval
$t_i = \frac{\pi}{2 \Delta B_{z,i}}$.
After time $t_1$ one obtains 
\begin{align}
 \left| \psi(t_1) \right\rangle =&{}\,
         \mathrm{e}^{-\frac{i}{2} \frac{\pi}{2}      \hat\sigma_x}
 \otimes \mathrm{e}^{-\frac{i}{2} \Delta B_{z,2} t_1 \hat\sigma_x} \left| S S \right\rangle
 \nonumber\\
 =&{}\, \frac{1}{\sqrt{2}}
        \Big(
                  \left| S   \right\rangle
              - i \left| T_0 \right\rangle
        \Big)
 \otimes \mathrm{e}^{ \frac{i}{2}
                      \frac{\pi}{2}
                      \frac{\Delta B_{z,2}}{\Delta B_{z,1}} \hat\sigma_x} \left| S \right\rangle \, .
\end{align}
Then,
the splitting $J_1$ is raised to suppress the rotation of qubit $1$ about $x$ axis,
while the rotation of qubit $2$ is being completed in time $\delta t= t_2-t_1$:
\begin{align}
 \left| \psi(t_1+\delta t = t_2) \right\rangle &=
         \mathrm{e}^{-\frac{i}{2} J_1\delta t             \hat\sigma_z}
 \otimes \mathrm{e}^{-\frac{i}{2} \Delta B_{z,2} \delta t \hat\sigma_x} \left| \psi(t_1) \right\rangle \nonumber\\
 &= \frac{1}{\sqrt{2}}
    \Big(
              \mathrm{e}^{-i \frac{J_1\delta t}{4}} \left| S   \right\rangle
          - i \mathrm{e}^{+i \frac{J_1\delta t}{4}} \left| T_0 \right\rangle
    \Big) \nonumber \\
 & \otimes \frac{1}{\sqrt{2}} 
 \Big(
           \left|S\right\rangle
       - i \left|T_0\right\rangle
 \Big) \, .
\end{align}
In order to remove the phases imprinted on qubits due to finite $\Delta B_{z,i}$, the spin echo sequences are carried out on each qubit. (Of course, in the more realistic setting in which $J$ fluctuates, the need to remove the influence of the slowest of these fluctuations on the final two-qubit state is a much stronger motivation to employ spin echo.)
The SE sequence consists of three steps:
the evolution with splitting $J_i$ over a chosen time interval,
$\pi$ rotation of a state about $x$ axis
(i.e.~$2t_i$ interval when the splitting $J_i$ is turned off),
which is followed by the evolution for the same time interval.
In this case, the durations of SE on each qubit are chosen so
that both sequences terminate at the same instant $\tau$.
Since $t_1 \neq t_2$ this requirement implies
that each SE starts at different time, and they last for unequal durations.
Simultaneously during the evolution intervals  excluding the periods of evolution corresponding to the $\pi$ rotations
both splittings are on
and hence
the two-qubit coupling $\hat H_\mathrm{int}$ is on as well,
thus allowing for qubits to entangle.
Figure \ref{fig:procedure} showcases the time dependence
of splittings $J_1$ and $J_2$ for the entirety of the entangling procedure.
At the end of the procedure, the following state is produced:
\begin{align}
 \left| \tau \right\rangle \equiv| \psi(\tau) \rangle 
&= 
\frac{1}{2}
   \mathrm{e}^{ -\frac{1}{2} i J_{12} (\tau-3t_2) } \left| S   S   \right\rangle  \nonumber\\
&+ \frac{i}{2}                                      \left| S   T_0 \right\rangle
 + \frac{i}{2}                                      \left| T_0 S   \right\rangle  \nonumber\\
&- \frac{1}{2}
   \mathrm{e}^{ -\frac{1}{2} i J_{12} (\tau-3t_2) } \left| T_0 T_0 \right\rangle \, .
 \label{eq:state_tau}
\end{align}
Recall that $t_2 > t_1$
because it was assumed
that $\Delta B_{z,1} \geqslant \Delta B_{z,2}$ without
loss of generality
(if $\Delta B_{z,2} > \Delta B_{z,1}$
simply relabel the qubits to obtain the same result). Note that in the currently considered case of fixed $J$ and $\Delta B_z$ the latter drops out from the final state $|\tau\rangle$.

It should be noted that for the discrete set of durations
$\tau = \tau_{\mathrm{ent}}^{(a)} = (2a-1) \frac{\pi}{J_{12}} + 3t_2$,
where $a$ is a natural number, 
the state $|\tau\rangle$ is maximally entangled,
specifically, for odd or even $a$ the entangling procedure generates the state
\begin{equation}
|\psi_\mathrm{o}\rangle = \frac{i}{2} \left( - |S   S  \rangle
                                    + |S   T_0\rangle
                                    + |T_0 S  \rangle
                                    + |T_0 T_0\rangle \right)
\label{eq:psi_odd}
\end{equation}
or
\begin{equation}
|\psi_\mathrm{e}\rangle = \frac{i}{2} \left(   |S   S  \rangle
                                    + |S   T_0\rangle
                                    + |T_0 S  \rangle
                                    - |T_0 T_0\rangle \right),
\label{eq:psi_even}
\end{equation}
respectively.
One can easily notice that the entangling procedure realizes a CPHASE gate:
$
-\mathrm{e}^{-i J_{12} \frac{\tau-3 t_2}{2}} | S   S   \rangle \langle T_0 T_0 |
-                                            | S   T_0 \rangle \langle T_0 S   |
-                                            | T_0 S   \rangle \langle S   T_0 |
-\mathrm{e}^{-i J_{12} \frac{\tau-3 t_2}{2}} | T_0 T_0 \rangle \langle S   S   |
$.

\section{The quantification of two-qubit entanglement} 
\label{sec:EntQuantif}                                 
The next step is
to quantify the level of entanglement
possessed by the two-qubit state $\left| \tau \right\rangle$
produced at the end of the entangling procedure.
For two-qubit states described by the density operator $\hat\rho$
the most commonly used measure of entanglement is the {\it concurrence}
\cite{Wootters_PRL98}.
It is defined as
\begin{equation}
 C(\hat\rho) = \mathrm{max}\left\{ 0, \lambda_1-\lambda_2-\lambda_3-\lambda_4\right\} \, ,
\end{equation}
where $\lambda_1 \geqslant \lambda_2 \geqslant \lambda_3 \geqslant \lambda_4$
are the square roots of the eigenvalues of the matrix $\hat \rho \hat{\tilde{\rho}}$, 
where $\hat{\tilde{\rho}} = (\hat\sigma_y \otimes \hat\sigma_y ) \hat \rho^* (\hat\sigma_y \otimes \hat\sigma_y )$.
Here ${\hat\rho}^*$ denotes the operation of complex conjugation
of each element of $\hat\rho$.
The concurrence ranges from $C=0$ for separable states
to $C=1$ for maximally entangled states.
The concurrence of a pure state is given by a simple formula:
$C = |\langle \psi | \hat\sigma_y \otimes \hat\sigma_y | \psi^*\rangle|$,
where $| \psi^*\rangle$ is a complex conjugation of $| \psi\rangle$.

For the family of output states
$\hat\rho(\tau) = | \tau \rangle \langle \tau |$
the concurrence is given by
\begin{equation}
 C\left( \hat\rho(\tau) \right) = \left| \sin \left[\frac{1}{2} J_{12}(\tau-3t_2) \right] \right|\,.
 \label{eq:Concurrence_ideal}
\end{equation}

One can make use of an alternative strategy to check
to what degree the resulting state is entangled:
high level (greater than 1/2) of fidelity
calculated between the actual state $|\tau\rangle$ and the expected entangled state
$| \psi_\mathrm{o} \rangle$ or $| \psi_\mathrm{e} \rangle$, defined as $F(|\tau\rangle, |\psi_\mathrm{o}\rangle) = |\langle\tau|\psi_\mathrm{o}\rangle|^2$, confirms the entanglement 
\cite{Shulman_Science12}.
In general, for mixed states, 
fidelity is defined as 
$F(\hat \rho(\tau), \hat\rho_\mathrm{o}) = \mathrm{Tr}\{\hat \rho(\tau) \hat\rho_\mathrm{o}\}$. 
If one of the states is pure, as we will be having below, fidelity is given by 
$F(\hat \rho(\tau), |\psi_\mathrm{o}\rangle) = \langle\psi_\mathrm{o}| \hat \rho (\tau)|\psi_\mathrm{o}\rangle$.

\begin{figure}[t]
\centering
 \includegraphics[width=\linewidth]{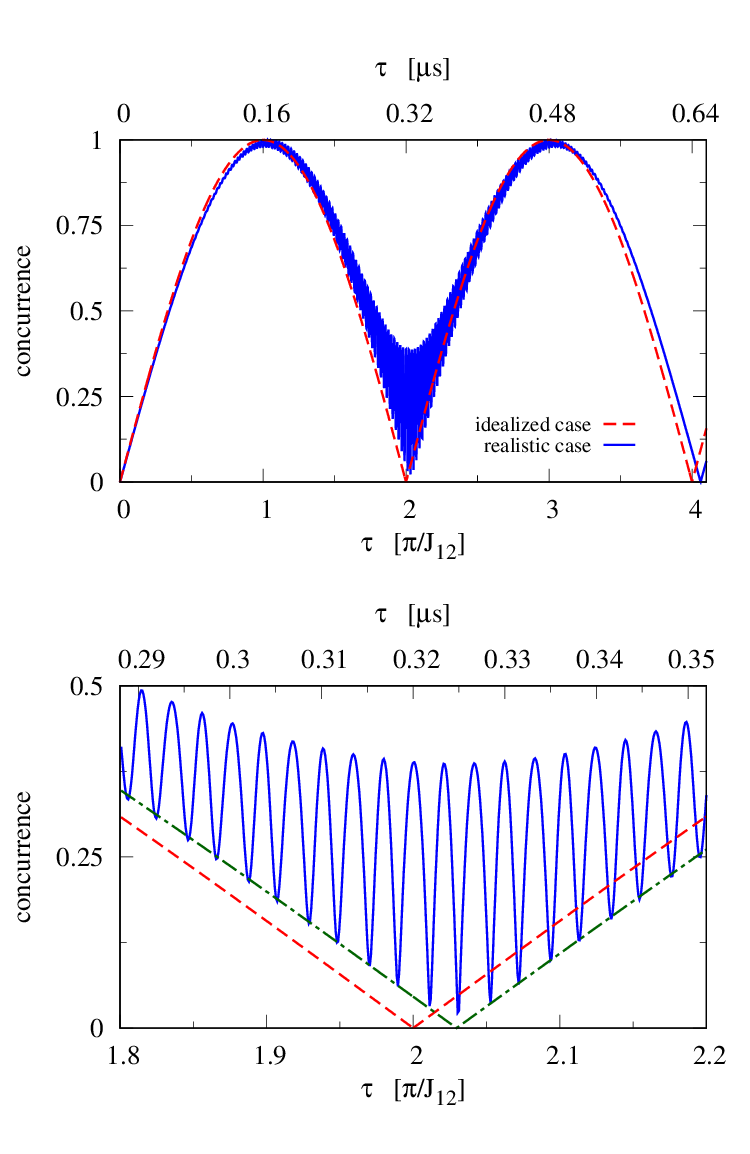}
 \caption{
           (Color online)~Top panel: Concurrence of the two-qubit state
           generated in the idealized realization of the entangling procedure described in Sec.~\ref{sec:2STq}
           ($\Delta B_{z,i}$ are on during the qubit rotations only,
           the rotations assumed to be perfect) plotted as red dashed line, and the same for a 
           more realistic procedure considered in Sec.~\ref{sec:EntQuantif}
           ($\Delta B_{z,i}$ are always on,
           albeit none of the parameters of the Hamiltonian are fluctuating) plotted as a solid blue line.
           The parameters are close to those from the experiment
           \cite{Shulman_Science12}:
           $J_1\! =\! J_2 \! = \! 1.2$ $\mu$eV (with corresponding frequency $J_i/h \! =\! 300$ MHz), $J_{12} \! =\! 1.29 \cdot 10^{-2}$ $\mu$eV, $\Delta B_{z,1}\! =\! \Delta B_{z,2} \! = 0.12$ $\mu$eV.
           Bottom panel: zoomed in region of the top panel in the vicinity of $\tau \approx 2 \frac{\pi}{J_{12}}$,
           red dotted line is the concurrence obtained in the idealized case,
           blue solid line is the concurrence obtained in a more realistic case 
           with constant parameters $J_i,~ J_{12}$, $\Delta B_{z, i}$,
           and green dash-dotted line is the concurrence
           obtained in the idealized case but with value of $J_{12}$ from the latter case.
         }
 \label{fig:C_ideal_const_param}
\end{figure}

In Fig.~\ref{fig:C_ideal_const_param} we show the results of numerical calculation of concurrence of the final two-qubit state as a function of procedure duration  $\tau$ in a model less idealized than in the previous section. While we are still assuming that $J_i$ and $\Delta B_{z,i}$ do not experience any fluctuations, we now keep $\Delta B_{z,i}$ fixed during the evolution (as it is the case in experiment), so that the evolution with finite $J_i$ does not amount to phase evolution in the computational basis of $S/T_{0}$ states of two qubits, thus necessitating numerical evaluation of the entanglement measure. 

Although the impact of always-on $\Delta B_{z,i}$ terms
amounts to a small drop of level of entanglement
of the resulting state compared to that from the idealized scenario described in the previous section (see Fig.~\ref{fig:C_ideal_const_param}), 
it reveals also another delicate effect: 
level of entanglement of the resulting state $|\tau\rangle$ 
starts to oscillate with frequency
of precession of single qubit states $\omega_i = \sqrt{\Delta B_{z,i}^2 + J_i^2}$ 
due to the fact
that in such conditions the qubits' states rotate around the axis
which does not coincide with $z$ axis exactly, 
but it is tilted in $xz$ plane because of presence $\Delta B_{z}$.
Characteristically,
the oscillations of entanglement of the state $\hat\rho(\tau)$
gradually increase their amplitude 
with increasing durations for $0 < \tau \lesssim 2 \frac{\pi}{J_{12}}$
and reach their maximum amplitude at $\tau \approx 2 \frac{\pi}{J_{12}}$.
Then the amplitude of oscillations decreases
for $2\frac{\pi}{J_{12}} \lesssim \tau \lesssim 4 \frac{\pi}{J_{12}}$.
The observed pattern of fast oscillation of two-qubit entanglement is periodic in duration $\tau$ of the entangling procedure with a period of about $4\frac{\pi}{J_{12}}$.
Such a pattern of oscillation amplitude $\tau$-dependence 
is a consequence of the entangling procedure design:
at $ \tau = 2 \frac{\pi}{J_{12}}$ 
the ideal resulting state 
$|\tau\rangle = -\frac{1}{2} \left(|S\rangle - i |T_0\rangle \right)^{\otimes 2}$
is unentangled 
due to a very specific combination of phases generated before and after the $\pi$ rotations
of qubits' states.
The two-qubit state 
that is produced in the middle of the idealized realization of the entangling procedure
(just before the step of $\pi$ rotations) 
is fully entangled.
However,
when the axes around which qubit states precess
are tilted from $z$ direction~\footnote{
For typical values of 
$\Delta B_{z, i} \approx 0.12$~$\mu$eV
and 
$J_i \approx 1.2$~$\mu$eV, 
the angle $\alpha$ between the $z$ axis and the actual axis of rotations is 
$\alpha = \arcsin \left(\Delta B_{z, i} / \sqrt{\Delta B_{z, i}^2 + J_i^2}\right) \approx 6^{\circ}$.
}, 
the initial superposition states of qubits do not rotate in the equatorial plane (as intended)
but in a slightly tilted plane.
After the $\pi$ rotation around $x$ axes 
those states land on the plane which is tilted off 
from the equatorial plane 
in the opposite  direction~\footnote{
Using a geometrical interpretation of qubit's state space,
one can notice that $\pi$ rotation about $x$ axis, 
$\hat R_x(\pi) =  \mathrm{diag}(1,-1,-1)$,
changes the sign of $z$ component of the qubit's Bloch vector.
Hence, for example, 
the maximally distant from the equatorial plane state 
$|\tilde x_-\rangle 
= \cos \left( \theta /2 \right) |S  \rangle
+ \sin \left( \theta /2 \right) |T_0\rangle$,
where 
$
\theta = 2 \arccos
\left( 
\sqrt{\Delta B_{z, i}^2 + J_i^2 - \Delta B_{z, i}\sqrt{\Delta B_{z, i}^2 + J_i^2}} 
\right.
$
$
\left.
/
\sqrt{2 \left(\Delta B_{z, i}^2 + J_i^2\right)}
\right)
> \frac{\pi}{2}
$,
with corresponding Bloch vector
$ \mathbf{r}_{|\tilde x_-\rangle}
= \cos \alpha \, \mathbf{i} - \sin \alpha \, \mathbf{k}
= \begin{pmatrix}
\cos \alpha \\
0\\
-\sin \alpha
\end{pmatrix}
$,
which is obtained from the initial state $|{-y}\rangle$
after a period 
$t = \pi / \left( 2 \sqrt{\Delta B_{z, i}^2 + J_i^2} \right)$
of the free evolution 
(the rotation by $\pi/2$ about an axis tilted from the $z$ direction with both present $\Delta B_{z, i}$ and $J_i$),
transforms into 
$\hat R_x(\pi) |\tilde x_-\rangle 
= |{\tilde x}_+\rangle 
$
with corresponding Bloch vector
$\mathbf{r}_{|{\tilde x}_+\rangle }
= \cos \alpha \, \mathbf{i} + \sin \alpha \, \mathbf{k}
= \begin{pmatrix}
\cos \alpha \\
0\\
\sin \alpha
\end{pmatrix}$, 
where $\mathbf{i}$, $\mathbf{k}$ are the unit vectors of $x$, $z$ directions, respectively.
The only qubit's states generated in the free evolution starting from $|{-y}\rangle$ 
which stay in the equatorial plane after application of $\hat R_x(\pi)$ are states
$|{\pm y}\rangle 
= \frac{1}{\sqrt 2} (|S\rangle \pm i |T_0\rangle)$
with corresponding Bloch vectors
$ \mathbf{r}_{|{\pm y}\rangle }
= \pm \mathbf{j}
= \begin{pmatrix}
0 \\
\pm 1\\
0
\end{pmatrix}$, 
where $\mathbf{j}$ is the unit vector of $y$ direction.
},
and subsequently the phases which qubits' states acquire  
during the second half of the procedure 
are no longer in a perfect correspondence to the previously obtained phases
and now they do not counterbalance each other,
so the final two-qubit state manifests unexpected entanglement.
One can also notice
that in the presence of constant $\Delta B_{z, i}$
the period of slow entanglement oscillations
is slightly longer compared to the idealized case
(see Fig.~\ref{fig:C_ideal_const_param}).
This fact cannot be illustrated with the help of Bloch sphere 
as it originates from a two-qubit interaction, 
but it is evident from the numerical diagonalization of the full two-qubit Hamiltonian:
when $\Delta B_{z, i}$ are present, 
the rate of acquisition of the desired two-qubit phase is slower
than that of the ideal case
(CPHASE gate is rotated from the basis of
$\{|S   S   \rangle,~|S   T_0 \rangle,~|T_0 S   \rangle,~|T_0 T_0 \rangle\}$ 
to the basis of eigenvectors of the full Hamiltonian,
and as a result the rate of two-qubit phase acquisition becomes lower).

\section{Entangling procedure in the presence of decoherence} 
\label{sec:EntProcedureDecoh}                                 
In the experiment 
\cite{Shulman_Science12} it turned out
that the two-qubit states obtained as outcome of the procedure
were indeed entangled
but only for short durations $\tau$ of the procedure.
Furthermore,
the maximal level of entanglement of the generated states was decreased
compared to that of the expected ones.
We consider below possible factors
that preclude one from obtaining the maximally entangled two-qubit states.

\subsection{Influence of fluctuations of magnetic field gradients 
            on efficiency of the entangling procedure             
           }                                                      
The magnetic field gradients between the QDs, which are the sources of finite $\Delta B_{z,i}$ (with $i=1$, $2$), are often produced 
by the polarized in an appropriate way nuclear spins
of the
atoms from which the sample is built
\cite{Foletti_NP09,Shulman_Science12}.
The spin bath is dynamically polarized before each iteration of the experiment
of entangling of qubits.
Due to the slowness of the intrinsic dynamics of the nuclear spin bath,
one does not expect any fluctuation on the time scale
of a single run of the experiment ($\sim 1~\mu \mathrm{s}$).
However,
possible variations of the values of $\Delta B_{z,i}$ 
from one run of the experiment to another
is the factor which can preclude from obtaining the maximally entangled states.
This effect is mainly caused by imprecise rotations of qubits' states around $x$ axis,
which are performed just after the initializations of the qubits in
$|S   S   \rangle$
state 
($\frac{\pi}{2}$ rotations)
and in the middle of the entangling procedure ($\pi$ rotations).
Such systematical errors lead to forming 
unequal superposition states of $| S \rangle$ and  $| T_0 \rangle$.
The influence of quasistatic fluctuations of $\Delta B_{z,i}$
on the efficiency of the entangling procedure
results in a decrease of the overall efficiency independently of $\tau$,
i.e.~the fluctuating quasistatically $\Delta B_{z,i}$
influences in a similar way the outcomes for all duration $\tau$
of the entangling procedure.

\begin{figure}[t]
\centering
\includegraphics[width=\linewidth]{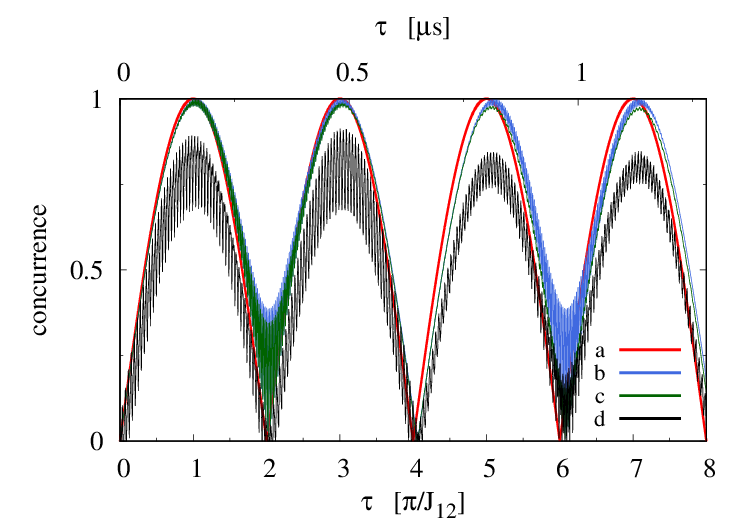}
\caption{
(Color online)~Concurrence of the two-qubit state~$\hat \rho(\tau)$
obtained in
(a)~idealized realization of entangling procedure (the same as red dashed line in Fig.~\ref{fig:C_ideal_const_param}),
(b)~entangling procedure with always-on constant
$\Delta B_{z,i}$ (the same as blue solid line in Fig.~\ref{fig:C_ideal_const_param}),
(c)~entangling procedure in which magnetic field gradients $\Delta B_{z,i}$ are constant during a single run of entangling procedure, but fluctuate quasistatically from one run to another (average over 10,000 realizations,
$\Delta B_{z,i} \sim \mathcal{N}(\overline{\Delta B_{z,i}},\, \sigma_{\Delta B_{z,i}})$,
$\sigma_{\Delta B_{z,i}} = 15\% \, \overline{\Delta B_{z,i}}$), here rotations around $x$ axis artificially kept perfect (i.e.~precisely $\frac{\pi}{2}$ at the beginning and $\pi$ in the middle of the procedure),
(d)~the~same as (c) but with imprecise rotations around $x$ axis which arise as a result of mismatch of the rotation time and the actual value of $\Delta B_{z,i}$.
All results are obtained under the assumption that $S$-$T_{0}$ splittings $J_i$ do not fluctuate and are switched on when needed (as shown in Fig.~\ref{fig:procedure}).
Values of parameters are the same as in Fig.~\ref{fig:C_ideal_const_param}.
}
 \label{fig:C_ideal_vs_DBz}
\end{figure}

This effect can be easily seen
when one considers the idealized realization of the procedure:
for simplicity, let us assume that only the first rotation was not exactly $\frac{\pi}{2}$ around $x$ axis
(errors of rotations  $\hat R_x(\frac{\pi}{2})$ and $\hat R_x(\pi)$ will accumulate -- there is no possibility that the next rotation cancels the error of the previous one). In such a case, the procedure will produce the state
$|\tau(\theta_1, \theta_2)\rangle =
  |S   S   \rangle (-i) \sin\frac{\theta_1}{2} \sin\frac{\theta_2}{2} \exp(-i\frac{\tau}{2} J_{12})
+ |S   T_0 \rangle (-i) \sin\frac{\theta_1}{2} \cos\frac{\theta_2}{2}
+ |T_0 S   \rangle (-i) \cos\frac{\theta_1}{2} \sin\frac{\theta_2}{2}
+ |T_0 T_0 \rangle      \cos\frac{\theta_1}{2} \cos\frac{\theta_2}{2} \exp(-i\frac{\tau}{2})
$,
where $\theta_i$ is the actual angle of rotation of the state of $i$th qubit around $x$ axis.
This state $|\tau(\theta_1, \theta_2)\rangle$ is maximally entangled
when superpositions of $|S\rangle$ and $|T_0\rangle$, created in each qubit from the separable two-qubit state $|SS\rangle$ after $\hat R_x(\frac{\pi}{2})$ rotation,
are equal (i.e.~all components have the same amplitudes):
at $\tau = \pi/J_{12}$ the state $|\tau(\theta_1, \theta_2)\rangle$ should become $|\psi_{\mathrm{o}}\rangle$, which is maximally entangled,
so one can estimate entanglement of $|\tau(\theta_1, \theta_2)\rangle$ by calculating 
fidelity
$F(|\tau(\theta_1, \theta_2)\rangle, |\psi_{\mathrm{o}}\rangle)
= |\langle\tau(\theta_1, \theta_2)|\psi_{\mathrm{o}}\rangle|^2
=\frac{1}{4} 
|   \sin\frac{\theta_1}{2}  \sin\frac{\theta_2}{2}
  + \sin\frac{\theta_1}{2}  \cos\frac{\theta_2}{2}
  + \cos\frac{\theta_1}{2}  \sin\frac{\theta_2}{2}
  + \cos\frac{\theta_1}{2}  \cos\frac{\theta_2}{2}
|^2
$, which has its maximum $F=1$ when $\theta_1 = \theta_2 = \frac{\pi}{2}$.
Any deviation of $\theta_1, \theta_2 \in[0,\pi]$ will reduce the degree of entanglement of the state $|\tau(\theta_1, \theta_2)\rangle$.
On the other hand, having a pure state $|\tau(\theta_1, \theta_2)\rangle$
it is possible to calculate analytically its concurrence
$C(|\tau(\theta_1, \theta_2)\rangle)
= |\langle\tau(\theta_1, \theta_2)| \hat \sigma_y \otimes \hat \sigma_y   |\tau^*(\theta_1, \theta_2)\rangle |
= 4 \sin\frac{\theta_1}{2}
    \cos\frac{\theta_1}{2}
    \sin\frac{\theta_2}{2}
    \cos\frac{\theta_2}{2}
    \sin(\frac{\tau}{2}J_{12})
$,
which, when $\tau = \pi/J_{12}$, varies from $C=1$ for $\theta_1=\theta_2=\frac{\pi}{2}$
to $C=0$ when $\theta_1,\theta_2  \rightarrow 0 \mathrm{~or~} \pi$.
Hence, the impact of quasistatical fluctuations of magnetic field gradients $\Delta B_{z,i}$ amounts to a loss of the maximal level of produced entanglement (see Fig.~\ref{fig:C_ideal_vs_DBz}).
This effect is independent of the duration $\tau$ of the procedure
and does not lead to a complete inability to yield some entaglement
when fluctuations of  $\Delta B_{z,i}$ are moderate or small.
Another important observation which comes from Fig.~\ref{fig:C_ideal_vs_DBz}
is that the presence of quasistatically fluctuating magnetic gradients $\Delta B_{z,i}$
during the entire entangling procedure,
does not degrade the efficiency of the procedure by a significant amount
while rotations of qubit states around $x$ axis are accurate 
(cf. line c, which is very close to line b,
with line d in Fig.~\ref{fig:C_ideal_vs_DBz}).

Consequently, in the rest of the paper, where we will focus on influence of  fluctuations of $J_i$ that will lead to complete decay of entanglement, we will neglect all the above-discussed effects of quasistatic fluctuations of $\Delta B_{z,i}$ and rotation errors caused by $\Delta B_{z,i}$ being finite, albeit small in comparison to $J_i$.

\subsection{Influence of fluctuations of exchange splittings on efficiency 
            of the entangling procedure                                    
           }                                                               
\label{sec:InfluenceJ}                             
\subsubsection{Quasistatic fluctuations of exchange splittings}
To begin with,
we consider the influence
of quasistatic fluctuations of exchange splittings $J_{i}$
on the entanglement generation.
The influence of quasistatically fluctuating exchange splittings $J_1,~J_2$
can be estimated by disregarding off-diagonal terms of the Hamiltonian
(but the qubit rotations involved in the entangling procedure are assumed to be perfect)
and performing the averaging of the density operator Eq.~(\ref{eq:DensityMatrix})
over the distribution of $J_1,~J_2$.
We assume that $J_1,~J_2$ fluctuate according to normal distribution
with mean values $\bar J_1,~\bar J_2$
and standard deviations $\sigma_{J_1},~\sigma_{J_2}$, respectively.

The idealized entangling procedure generates states which are described by the following density operator:
\begin{align}
 \hat \rho (\tau) &= | \psi(\tau) \rangle \langle \psi(\tau) | \nonumber \\
 &= \frac{1}{4}
 \begin{pmatrix}
  1              &
 -i \phi(\tau)   &
 -i \phi(\tau)   &
 -1                \\
  i \phi^*(\tau) &
  1              &
  1              &
 -i \phi^*(\tau)   \\
  i \phi^*(\tau) &
  1              &
  1              &
 -i \phi^*(\tau)   \\
 -1              &
  i \phi(\tau)   &
  i \phi(\tau)   &
  1
 \end{pmatrix},
 \label{eq:DensityMatrix}
\end{align}
where $\phi(\tau) = \exp \left( -i \frac{\tau}{2} J_{12} \right)$. After averaging over quasistatic fluctuations of $J_1,~J_2$ we obtain the density operator
\begin{align}
 \langle \hat \rho (\tau) \rangle = \frac{1}{4}
 \begin{pmatrix}
  1                            &
 -i \langle\phi(\tau)\rangle   &
 -i \langle\phi(\tau)\rangle   &
 -1                              \\
  i \langle\phi^*(\tau)\rangle &
  1                            &
  1                            &
 -i \langle\phi^*(\tau)\rangle   \\
  i \langle\phi^*(\tau)\rangle &
  1                            &
  1                            &
 -i \langle\phi^*(\tau)\rangle   \\
 -1                            &
  i \langle\phi(\tau)\rangle   &
  i \langle\phi(\tau)\rangle   &
  1
 \end{pmatrix},
 \label{eq:DensityMatrix_QSAverage}
\end{align}
where 
\begin{align}
 \langle \phi(\tau) \rangle 
 &= \frac{2K}{\sqrt{4K^2 + \sigma_{J_1}^2 \sigma_{J_2}^2 \tau^2}}
  \exp\left(- i\frac{ 4 \bar J_1 \bar J_2 K \tau} {8K^2 + 2\sigma_{J_1}^2 \sigma_{J_2}^2 \tau^2}  \right) \nonumber \\
 &\times \exp\left( -\frac{(\bar J_1^2 \sigma_{J_2}^2 + \bar J_2^2 \sigma_{J_1}^2) \tau^2}{8K^2 + 2\sigma_{J_1}^2 \sigma_{J_2}^2 \tau^2} \right),
 \label{eq:coherence_QS}
\end{align}
where 
constant
$K~=~{\bar J_1 \bar J_2} / {\bar J_{12}}$,
and $\bar J_{12} = \frac{\pi}{\tau_{\mathrm{ent}}}$.
In the experiment
\cite{Shulman_Science12}
the values of parameters were as follows:
$\bar J_1 \! =\! 1.16$ $\mu$eV, $\bar J_2 \! =\! 1.32$ $\mu$eV, $\bar J_{12} \! =\! 1.29 \cdot 10^{-2}$ $\mu$eV, $\overline{\Delta B_{z,1}}\! =\! \overline{\Delta B_{z,2}} \! = 0.12$ $\mu$eV
(so $t_2 = 0$ in the Eq.~(\ref{eq:Concurrence_ideal})).
Note that we are using $J_{12}$ twice larger than the value reported in Ref.~\cite{Shulman_Science12}. However, with this value we obtain the period of oscillations of concurrence in agreement with experimental data, i.e.~the first maximum of entanglement occurs at $\tau \! =\! \pi/J_{12} \! \approx 160$ ns. 

Entanglement of Eq.~(\ref{eq:DensityMatrix_QSAverage})
as a function of duration $\tau$ is shown
in the top panel of Fig.~\ref{fig:C_J_noises}
for $J_i$ drawn from normal distribution
with standard deviations $\sigma_{i} = 15\% \bar{J}_{i}$.
Due to quasistatic fluctuations of $J_{i}$ the overall efficiency
of the entangling procedure decreases with increase of its duration $\tau$.
Although the direct impact of the quasistatic fluctuations of $J_1$, $J_2$
on the resulting two-qubit state is completely removed
by utilizing simultaneous Hahn echo sequence on each qubit,
the entangling interaction between qubits,
which is determined by two-qubit interaction energy $J_{12} \propto J_1 J_2$, 
remains sensitive to the fluctuations,
and this causes decay of the entangling procedure efficiency
with increasing duration $\tau$.

\subsubsection{Dynamical fluctuations of exchange splittings}

In the experiment 
[\onlinecite{Shulman_Science12}]
the procedure of entangling two $S$-$T_{0}$ qubits was based on the SE procedure.
While the SE perfectly cancels the impact of quasistatic single-qubit noises on the end state,
in the case of dynamical fluctuations
it helps to refocus the state of the qubits only partially.
Moreover, the two-qubit interaction part of the evolution operator
that describes the entangling procedure
(which is responsible for the entanglement generation)
is not affected by the SE procedure, and consequently it is susceptible to noisy electric fields (leading to noise in $J_{i}$ and $J_{12}$) 
in the same way as in FID experiment.
As a result, the overall efficiency of the entangling procedure decays
with increasing its duration $\tau$.

In order to estimate analytically the influence of dynamical fluctuations of exchange splittings,
we approximate the Hamiltonian
of the system by its diagonal
neglecting the off-diagonal terms associated with magnetic field gradients $\Delta B_{z,i}$,
which were an order of magnitude smaller than exchange splittings $J_{i}$ in the experiment~[\onlinecite{Shulman_Science12}]:
\begin{align}
 \hat H_{\mathrm{2q}} \approx \hat H_{\mathrm{2q}}^{\mathrm{diag}} &= \frac{1}{2} \Big(  J_1(t) \hat \sigma_z   \otimes \mathds{1}
                                          + J_2(t) \mathds{1} \otimes \hat \sigma_z
 \nonumber \\ 
             &+ \frac{1}{2} J_{12}(t) (\hat \sigma_z + \mathds{1}) \otimes (\hat \sigma_z + \mathds{1}) \Big).
\end{align}

Assuming perfect rotations of qubits' states,
averaged density operator elements of the resulting two-qubit state
after performing the entangling procedure are
\begin{align}
 &\left\langle \rho_{ab,cd}(\tau) \right\rangle =
 \left\langle \langle a b | \hat \rho(\tau) | c d \rangle \right\rangle \nonumber \\
 &=   \langle \langle a b |
 \hat U_{\mathrm{SE}}(\tau) \hat \rho^{\mathrm{initial}} \hat U_{\mathrm{SE}}^{\dagger}(\tau)
                                            | c d \rangle \rangle \nonumber \\
 &= \langle \langle ab | \hat U_{\mathrm{FID}} \Big(\frac{\tau}{2},\tau\Big) \Big((-i \sigma_x)\otimes(-i \sigma_x)\Big) \hat U_{\mathrm{FID}} \Big(0,\frac{\tau}{2}\Big) \nonumber \\
 & \times \sum_{kl,mn} \rho_{kl,mn}^{\mathrm{initial}} |kl\rangle \langle mn| \nonumber \\
 & \times \hat U_{\mathrm{FID}}^{\dagger} \Big(0,\frac{\tau}{2}\Big) \Big((i\sigma_x) \otimes (i\sigma_x)\Big) \hat U_{\mathrm{FID}}^{\dagger} \Big(\frac{\tau}{2},\tau\Big) |cd\rangle \rangle,
\end{align}
where evolution operator 
$\hat U_{\mathrm{FID}}(t_1,t_2) = \exp\left({-i \int_{t_1}^{t_2} \hat H_{\mathrm{2q}}(t) \mathrm{d}t}\right)
\approx \exp\left({-i \int_{t_1}^{t_2} \hat H_{\mathrm{2q}}^{\mathrm{diag}}(t) \mathrm{d}t}\right)$.

Analyzing the two-qubit system,
we consider two distinct possibilities of dynamical fluctuations:
splitting energies $J_1$, $J_2$ could fluctuate independently, i.e.~$J_i(t) = \bar J_i + \delta J_i(t)$,
or their fluctuations may have a common source
$J_i(t) = \bar J_i + s_i \delta J(t)$,
where $s_i \in [ 0, 1 ]$  is a coupling of $i$th qubit to the noise. Note that correlations of low-frequency charge noises affecting two quantum dots separated by $\sim \! 100$ nm distance have been observed in experiments 
\cite{Boter_PRB20,Yoneda_arXiv22}.
Correspondingly, the two-qubit coupling in the former case reads
\begin{align}
J_{12}(t) &= \frac{J_1(t) J_2(t)}{K} =
\frac{1}{K} \left[  \bar J_1 + \delta J_1(t) \right]
            \left[  \bar J_2 + \delta J_2(t) \right]
\nonumber\\
            & \approx
\frac{1}{K} \left[  \bar J_1 \bar J_2
             + \bar J_2 \delta J_1(t)
             + \bar J_1 \delta J_2(t) \right] \,\, ,
\end{align}
and in the latter case 
\begin{align}
J_{12}(t) & = \frac{J_1(t) J_2(t)}{K} =
\frac{1}{K} \left[ \bar J_1 + s_1 \delta J(t) \right]
            \left[ \bar J_2 + s_2 \delta J(t) \right]
\nonumber\\
            & \approx
\frac{1}{K} \left[  \bar J_1 \bar J_2
             + \left(s_1 \bar J_1
             +       s_2 \bar J_2 \right) \delta J(t) \right] \,\, .
  \end{align}
Note that we neglect here the quadratic in noises terms $\propto \delta J_1(t) \delta J_2(t)$ and $\propto (\delta J(t))^2$ .

For the case of independent (completely uncorrelated) charge noises that affect $J_1(t)$ and $J_2(t)$,
the average density operator elements of the generated state are
\begin{align}
 &\langle \rho_{ab,cd}(\tau) \rangle =
 \langle \langle a b | \hat \rho(\tau) | c d \rangle \rangle \nonumber \\
 &=\rho_{-a-b,-c-d}^{\mathrm{initial}} \mathrm{e}^{ -i \frac{\bar J_1 \bar J_2}{4 K}(ab-cd) \tau } \nonumber \\
 &\times \left\langle \exp \Big[ - \frac{i}{2} \Big( \frac{\bar J_2}{2 K} (ab-cd) \int_0^{\tau} \mathrm{d}t \, \delta J_1 (t) f_{\mathrm{FID}}(t) \right. \nonumber \\ 
 &\left. + \big( c-a + \frac{\bar J_2}{2 K}  (c+d-a-b) \big) \int_0^{\tau} \mathrm{d}t \, \delta J_1(t) f_{\mathrm{SE}}(t) \Big) \Big]  \right\rangle \nonumber \\ 
 &\times \left\langle \exp \Big[ - \frac{i}{2} \Big( \frac{\bar J_1}{2 K} (ab-cd) \int_0^{\tau} \mathrm{d}t \, \delta J_2 (t) f_{\mathrm{FID}}(t) \right. \nonumber \\ 
 &\left. + \big( d-b + \frac{\bar J_1}{2 K}  (c+d-a-b) \big) \int_0^{\tau} \mathrm{d}t \, \delta J_2(t) f_{\mathrm{SE}}(t) \Big) \Big]  \right\rangle
\end{align}
\begin{align}
 &=\rho_{-a-b,-c-d}^{\mathrm{initial}} \mathrm{e}^{ -i \frac{\bar J_1 \bar J_2}{4 K}(ab-cd) \tau } \nonumber \\
 &\times \exp \Big[ -\frac{1}{2} \Big\{ \Big( \frac{\bar J_2}{4 K}(ab-cd) \Big)^2 \chi_{\mathrm{FID},J_1}(\tau) \nonumber \\
 &+ \Big( \frac{1}{2}\big( c-a + \frac{\bar J_2}{2 K}(c+d-a-b) \big) \Big)^2 \chi_{\mathrm{SE},J_1}(\tau) \Big\} \Big] \nonumber \\
 &\times \exp \Big[ -\frac{1}{2} \Big\{ \Big( \frac{\bar J_1}{4 K}(ab-cd) \Big)^2 \chi_{\mathrm{FID},J_2}(\tau) \nonumber \\
 &+ \Big( \frac{1}{2}\big( d-b + \frac{\bar J_1}{2 K}(c+d-a-b) \big) \Big)^2 \chi_{\mathrm{SE},J_2}(\tau) \Big\} \Big],
 \label{eq:DensityMatrix_UnCorrNoises}
\end{align}
where $a,\, b, \, c, \,d \,\in \, \{1,-1\}$
(these parameters code the
basis $\{ |SS\rangle, |ST_0\rangle, |T_0S\rangle, |T_0T_0\rangle \}$
as $\{ {11}, {1-\!1}, {-11}, {-1-\!1} \}$ in indices of density operator elements),
and $\chi_{\mathrm{FID}, J_i}(\tau),~ \chi_{\mathrm{SE}, J_i}{(\tau)}$ are the attenuation factors that describe the influence of dynamical noise of $J_i(t)$
in the case of FID (with constant time domain filter function $f_{\mathrm{FID}}(t)$)
or SE (with time domain filter function with a single sign inversion $f_{\mathrm{SE}}(t)$).
For noise spectrum of the form $S(\omega) \! =\! A/\omega^\beta$ they are given by
\begin{align}
\chi_{\mathrm{FID}}(\tau) &= \frac{4A} {\pi} \int_{0}^{\infty} \frac{\mathrm{d} \omega}{\omega^{2+\beta}}
                                                      \sin^2 \frac{\omega \tau}{2} \,\, , \\ 
\chi_{\mathrm{SE}}(\tau)  &= \frac{16A}{\pi} \int_{0}^{\infty} \frac{\mathrm{d} \omega}{\omega^{2+\beta}}
                                                      \sin^4 \frac{\omega \tau}{4} \,\, ,
\end{align}
see Appendix \ref{app:chi} for the details of derivation and simple analytical approximations in considered here cases of $\beta \! =\! 0.7$ and $2$. 

\begin{figure}[t]
\centering
  \includegraphics[width=0.776\linewidth]{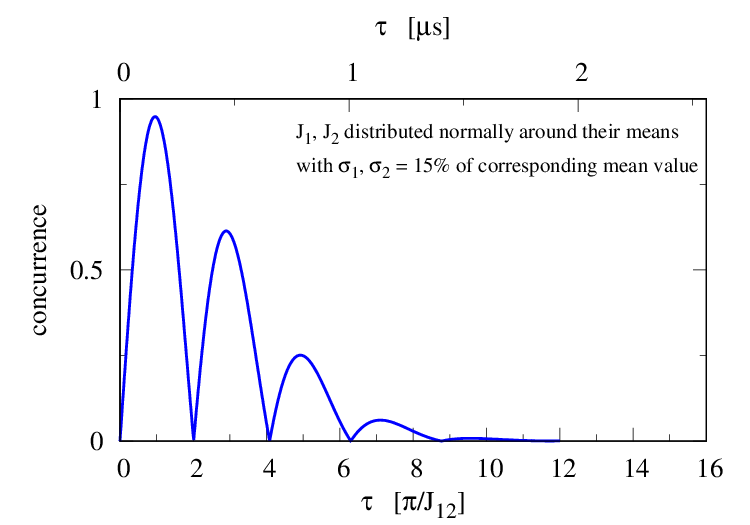}
  \includegraphics[width=0.776\linewidth]{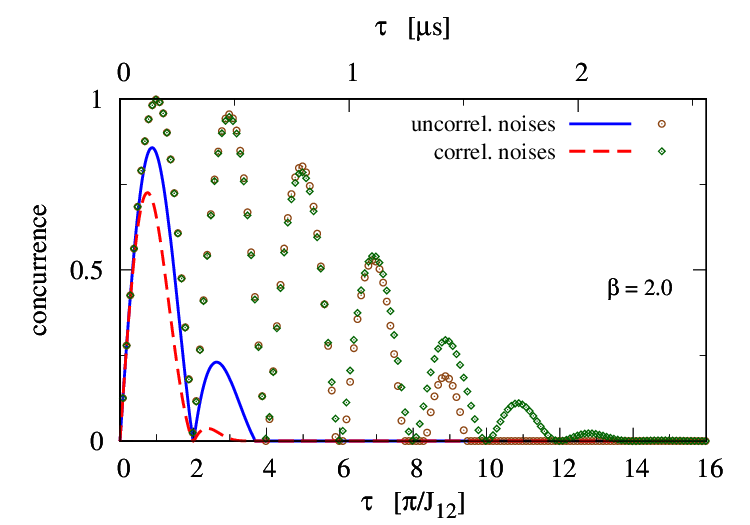}
  \includegraphics[width=0.776\linewidth]{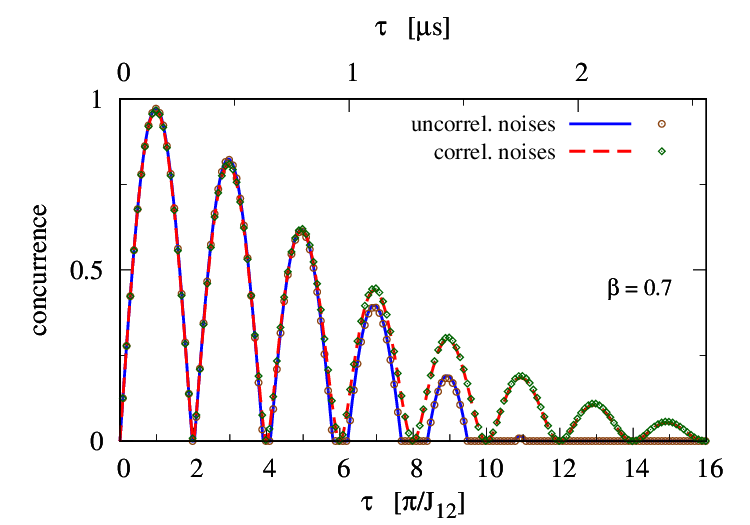}
  \caption{
            (Color online)
            Concurrence of the two-qubit state $\langle\hat \rho(\tau)\rangle$
            as a function of the duration $\tau$ of the entanglement generation procedure.
            Lines are results for the case in which fluctuations of $J_i$ and $J_{12}$ are taken into account, while results in which the fluctuations of $J_{12}$ were artificially turned off are shown with open symbols.
            Top panel:  The case of quasistatically fluctuating  $J_1$, $J_2$ 
            with standard deviations $\sigma_{i} = 0.15 \bar{J}_{i}$ (Eq.~(\ref{eq:DensityMatrix_QSAverage})).
            Middle and bottom panels: The cases of dynamically fluctuating $J_1$, $J_2$, and hence $J_{12}$, 
            due to $1/f^{\beta}$ noise that is uncorrelated (blue solid line) or perfectly correlated (red dashed lines) for the two qubits.
            The power of the noise affecting each qubit
            was chosen to be such
            that ensures the time scale of decay of single-qubit SE signal
            like in the experiment~\cite{Dial_PRL13}:
            $S$-$T_{0}$ qubit having $J \!= \! 1.16$~$\mu$eV shows
            $T_{\mathrm{SE}} \approx 1.6~\mu s$,
            $S$-$T_{0}$ qubit having $J \!= \! 1.32$~$\mu$eV shows
            $T_{\mathrm{SE}} \approx 1.4~\mu s$.
          }
  \label{fig:C_J_noises}
\end{figure}

Owing to the fact that the approximated Hamiltonian is diagonal,
the density operator undergoes the decoherence of pure dephasing type.
There are two essentially distinct types of off-diagonal elements
of two-qubit density operator. 
The density operator elements with a single spin flip:
$\langle \rho_{ 1 1, 1-1}(\tau) \rangle$,
$\langle \rho_{ 1 1,-1 1}(\tau) \rangle$,
$\langle \rho_{ 1-1,-1-1}(\tau) \rangle$,
$\langle \rho_{-1 1,-1-1}(\tau) \rangle$
and their Hermitian conjugated partners
diminish mainly due to decrease of single-qubits' signals
$\propto \exp \left( -\frac{1}{2} \chi_{\mathrm{SE}} (\tau) \right)$,
whereas elements with two spin flips:
$\langle \rho_{ 1 1,-1-1}(\tau) \rangle$,
$\langle \rho_{ 1-1,-1 1}(\tau) \rangle$
and their Hermitian conjugated partners
decay two times faster as both qubits make their contribution to the decay
$\propto \exp \left( -\chi_{\mathrm{SE}} (\tau) \right)$.
Hence,
the scale on which one can expect the generation of entangled state
is limited from above by the single-qubit SE decay time.

In the case of correlated noises,
$J_i(t) = \bar J_i + s_i \delta J(t)$,
the averaged density operator elements
are of the following form:
\begin{align}
 &\langle \rho_{ab,cd}(\tau) \rangle =
  \langle \langle a b | \hat \rho(\tau) | c d \rangle \rangle \nonumber \\
 &=\rho_{-a-b,-c-d}^{\mathrm{initial}} \mathrm{e}^{ -i \frac{\bar J_1 \bar J_2}{4K}(ab-cd) \tau } \nonumber \\
 &\times \left\langle \exp \Big[ - \frac{i}{2} \Big\{ \frac{s_1 \bar J_2 + s_2 \bar J_1}{2K} (ab-cd) \right. \nonumber \\
 &\times \int_0^{\tau} \mathrm{d}t \, \delta J (t) f_{\mathrm{FID}}(t)  \nonumber \\
 & + \Big( s_1 \big( c-a + \frac{\bar J_2}{2 K}(c+d-a-b) \big) \nonumber \\
 &        + s_2\big( d-b + \frac{\bar J_1}{2 K}(c+d-a-b) \big) \Big) \nonumber \\
 &\left. \times \int_0^{\tau} \mathrm{d}t \, \delta J(t) f_{\mathrm{SE}}(t) \Big\}  \Big] \right\rangle
\end{align}
\begin{align} 
 &=\rho_{-a-b,-c-d}^{\mathrm{initial}} \mathrm{e}^{ -i \frac{\bar J_1 \bar J_2}{4K}(ab-cd) \tau } \nonumber \\
 &\times \exp \Big[ - \frac{1}{2} \Big\{  \Big( \frac{s_1 \bar J_2 + s_2 \bar J_1}{4K} (ab-cd) \Big)^2 \chi_{\mathrm{FID}}(\tau)  \nonumber \\
 & + \Bigg( \frac{1}{2} \Big( s_1 \big( c-a + \frac{\bar J_2}{2 K}(c+d-a-b) \big) \nonumber \\
 &                          + s_2 \big( d-b + \frac{\bar J_1}{2 K}(c+d-a-b) \big)  \Big)  \Bigg)^2 \chi_{\mathrm{SE}}(\tau) \Big\} \Big] .
 \label{eq:DensityMatrix_CorrNoise}
\end{align}

The key qualitative feature of Eqs.~(\ref{eq:DensityMatrix_UnCorrNoises}) and (\ref{eq:DensityMatrix_CorrNoise}) is the presence of terms  proportional to $\chi_{\mathrm{SE}}$, in which the low-frequency noise is suppressed by the echo procedure, and of terms proportional to $\chi_{\mathrm{FID}}$, related to fluctuations of interqubit interaction, in which the low-frequency noise spectrum fully contributes to dephasing.

In Fig.~\ref{fig:C_J_noises}
the amount of entanglement is presented in the case
of uncorrelated noises for two exponents characterizing $1/f^\beta$  noise,
$\beta = 2.0$ and $0.7$ (middle and bottom panels, respectively).
As can be seen in the bottom panel of Fig.~\ref{fig:C_J_noises},
for $\beta \! < \! 1$
the decay of the overall efficiency of the entangling procedure is mainly caused
by influence of fluctuations of splittings of individual qubits (which also fully determines the decay of the fidelity of single-qubit coherence). 
On the other hand,
for $\beta > 1$,
(e.g.~$\beta=2.0$, the middle panel of Fig.~\ref{fig:C_J_noises}), the decay of the overall entanglement generation efficiency
is mostly due to the infidelity of the entangling gate,
which is realized by dynamically fluctuating two-qubit term. This is due to the fact that for noise that is very strongly concentrated at lowest frequencies, single-qubit noise is very efficiently suppressed by the echo procedure, and the non-echoed fluctuations of two-qubit interaction, which are the sources of terms $\propto \chi_{\mathrm{FID}}/K$ in the above expressions for two-qubit coherences, are dominating the dephasing of the final state.
Thus, the effect of dynamical noise of $J_i$ with high value of $\beta$ on the final state
is qualitatively the same as that of quasistatic fluctuations of splittings $J_i$ (see the top panel of Fig.~\ref{fig:C_J_noises})),
where single-qubit terms cancel out perfectly (thanks to applying Hahn echo on each qubit)
and the fidelity of two-qubit entangling gate,
which is susceptible to fluctuations as in FID experiment,
is diminishing when the duration $\tau$ of the procedure becomes longer.

\begin{figure}[t]
\centering
\includegraphics[width=0.622\linewidth]{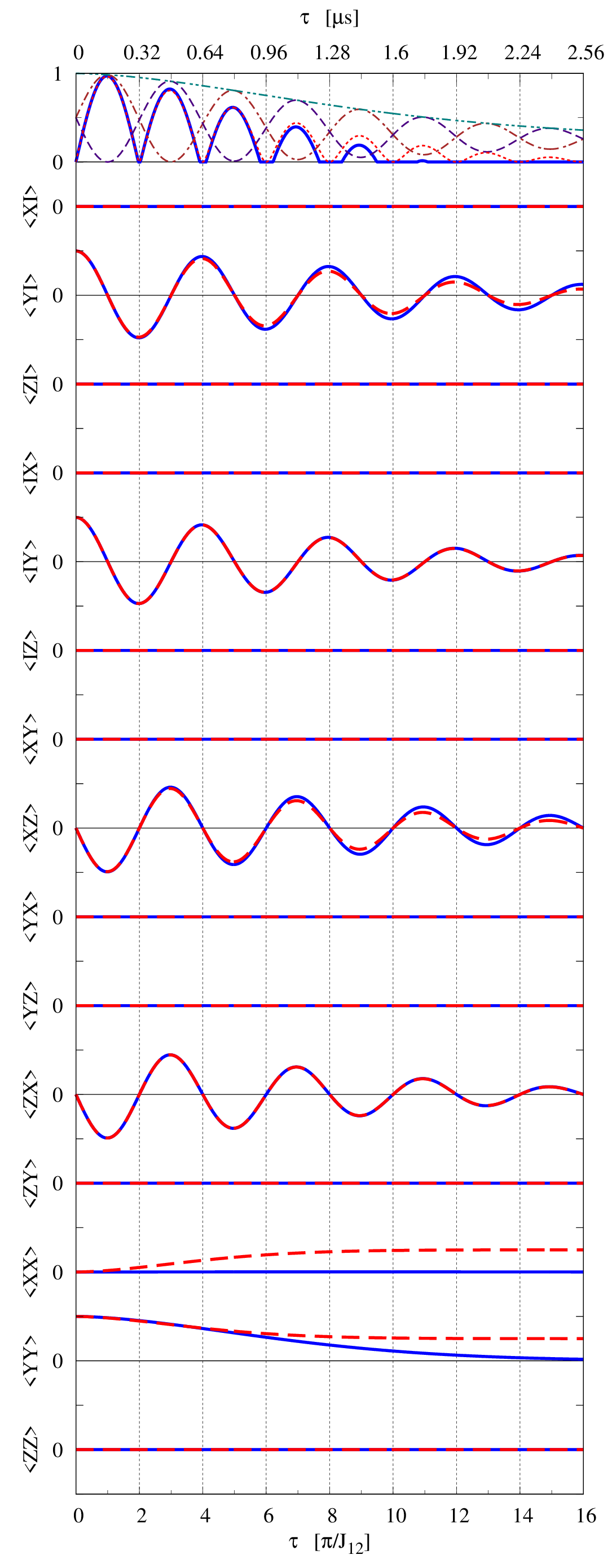}
\caption{
          (Color online)~Pauli set for the case of $1/f^{0.7}$ noise.
          Blue lines show calculated Pauli set for two independent noises
          of $J_1(t), J_2(t)$,
          red lines show calculated Pauli set for correlated noises
          $J_1(t)=J_2(t)$.
          In the top panel for the case of independent noises the following measures are shown:
          blue solid line is concurrence of $\langle\hat\rho(\tau)\rangle$,
          brown dashed line is fidelity
          $\langle \psi_{\mathrm{o}} | \langle\hat \rho(\tau)\rangle | \psi_{\mathrm{o}} \rangle$ (see Eq.~(\ref{eq:psi_odd})),
          purple dash-dotted line is fidelity
          $\langle \psi_{\mathrm{e}} | \langle\hat \rho(\tau)\rangle | \psi_{\mathrm{e}} \rangle $ (see Eq.~(\ref{eq:psi_even})),
          turquoise dotted line is fidelity
          $\langle \psi(\tau) | \langle\hat \rho(\tau)\rangle | \psi(\tau) \rangle$ (see Eq.~(\ref{eq:state_tau}); 
          and red dotted line is concurrence calculated for correlated noises $J_1(t) = J_2(t)$.
          The power of the noise affecting each qubit was chosen to be such
          that ensures the time scale of decay of single-qubit SE signal
          like in the experiment
          \cite{Dial_PRL13}:
          $S$-$T_{0}$ qubit having $J \!= \! 1.16$~$\mu$eV shows
          $T_{\mathrm{SE}} \approx 1.6~\mu s$,
          $S$-$T_{0}$ qubit having $J \!= \! 1.32$~$\mu$eV shows
          $T_{\mathrm{SE}} \approx 1.4~\mu s$.
          In the case of fully correlated noises $J_1(t) = J_2(t)$,
          the higher power of the noise has been chosen for both qubits.
        }
 \label{fig:PauliSet_07}
\end{figure}

In Fig.~\ref{fig:PauliSet_07}
the two Pauli sets for two-qubit states created in the entangling procedure
are presented
for the case of two uncorrelated $1/f^{0.7}$ noises (blue lines)
and for the case of two fully correlated $1/f^{0.7}$ noises (red lines).
In noise-free experiment, one expects
that the only nonzero two-qubit correlations
$\langle \hat \sigma_i \otimes \hat \sigma_j \rangle =
 \mathrm{Tr} \{ \left( \hat \sigma_i \otimes \hat \sigma_j \right) \hat \rho(\tau)  \}$
are 
\begin{align}
 \langle \hat \sigma_x   \otimes \hat \sigma_z   \rangle &=
 \langle \hat \sigma_z   \otimes \hat \sigma_x   \rangle = -\sin J_{12} \frac{\tau}{2}, \\
 \langle \hat \sigma_z   \otimes \mathds{1} \rangle &=
 \langle \mathds{1} \otimes \hat \sigma_y   \rangle =  \cos J_{12} \frac{\tau}{2}, \\
 \langle \hat \sigma_y   \otimes \hat \sigma_y   \rangle &= 1.
\end{align}
Dynamical fluctuations of splittings $J_i$ destroy these correlations
and diminish their amplitude with increasing duration $\tau$.
It is important to notice
that fully correlated noises always lead
to decreased but nonzero value
of $\langle \hat \sigma_y \otimes \hat \sigma_y \rangle \stackrel{\tau \rightarrow \infty}{=}
2\Re \rho_{1 -1, -1 1}^{\mathrm{initial}} = \frac{1}{2}$,
and at the same time new two-qubit correlation
$\langle \hat \sigma_x \otimes \hat \sigma_x \rangle \stackrel{\tau \rightarrow \infty}{=}
2\Re \rho_{1 -1, -1 1}^{\mathrm{initial}} = \frac{1}{2}$
is generated.
Therefore, the spatial correlations of noises can have a visible impact
on the evolution of resulting state $\hat \rho(\tau)$
and components of Pauli set.
Hence,
one can make use of this fact
to estimate to what degree the noises were correlated in the experiment.
By comparison of experimental data
(Fig.~3 in Ref.~[\onlinecite{Shulman_Science12}])
with simulated results (Fig.~\ref{fig:PauliSet_07}) 
one may deduce
that in the experiment~[\onlinecite{Shulman_Science12}]
noises of splittings $J_1$ and $J_2$ were uncorrelated.

\section{Conclusions}    
\label{sec:Conclusions}  
We have theoretically analyzed the creation and evolution of entanglement of two double quantum dot-based $S$-$T_{0}$ qubits measured in Ref.~\cite{Shulman_Science12} while taking into account realistic charge and nuclear noise affecting the qubits.
We have confirmed that 
it is possible to have nearly maximal coherence signal of a single $S$-$T_{0}$ qubit
in the presence of quasistatic fluctuations of either exchange splitting $J$
or magnetic-field gradient $\Delta B_{z}$
by performing spin echo procedure on the qubit.
Then, we have shown 
that in the system of two $S$-$T_{0}$ qubits
quasistatic fluctuations of $\Delta B_{z, i}$ lead only to partial decrease
of overall efficiency of the entangling procedure due to imprecise rotations of the qubits' states.

Both quasistatic or dynamical fluctuations of exchange splittings $J_1,~J_2$,
and two-qubit coupling $J_{12} \propto J_1 J_2$ lead to decay of overall efficiency
of the entangling procedure with increasing its duration $\tau$.
The level of correlation of charge noises,
as well as their exact functional form (i.e.~value of parameter $\beta$ characterizing the $1/f^\beta$ noise affecting $J$)
translates in a distinctive manner into the shape of decay of two-qubit entanglement
as a function of procedure duration $\tau$.
Decay of the overall efficiency of entangling procedure
may arise as a result of the infidelity of single qubit operations
(due to dynamical fluctuations of splittings $J_1,~J_2$)
or may be caused by infidelity of entangling gate
(due to fluctuations of two-qubit coupling $J_{12}$).

Comparison of experimental data from Ref.~[\onlinecite{Shulman_Science12}] with our calculations shows
that the charge noises in the system of two $S$-$T_{0}$ qubit investigated there were uncorrelated.
The main reason of the decreased level of entanglement of the resulting two-qubit state
is infidelity of single-qubit operations (as we have obtained for $1/f^\beta$ noise with $\beta \! = \! 0.7$ consistent with noise observed in other experiments
\cite{Dial_PRL13} on samples similar to those used in Ref.~\cite{Shulman_Science12}),
whereas contribution of the non-ideal two-qubit gate is negligible in the considered entangling procedure
in the regime $\Delta B_{z, i} \ll J_i$. We predict that for $J$ noises of more prominently low-frequency character (i.e.~$1/f^\beta$ with $\beta$ closer to $2$ than $1$), the fluctuations of the two-qubit interactions, which are not echoed by $\pi$ pulses applied to the two qubits separately, will become the main factor suppressing the maximal entanglement achievable in the considered procedure. We have also identified qualitative features of long-time behavior of two-qubit observables from the Pauli set that should be visible when the exchange splitting noises for the two qubits are correlated.


\section*{Acknowledgements}   
\label{sec:Acknowledgements}  
Authors thank Piotr Sza{\'n}kowski for helpful discussions.
This work was supported by Polish National Science Centre (NCN),
grant no.~DEC-2012/07/B/ST3/03616.

\pagebreak
\newpage


\appendix
\label{sec:Appendix}  
\section{Components of a Single $S$-$T_{0}$ Qubit
as Functions of Duration $\tau$
} \label{app:single}
In the ideal case, FID signals
(i.e.~average $\langle \hat \sigma_i^{\mathrm{FID}}(\tau) \rangle$ components  of $S$-$T_{0}$ qubit)
evolve as follows.
\begin{align}
 &\langle \hat \sigma_x^{\mathrm{FID}}(\tau) \rangle
 = \langle {-y}|   \mathrm{e}^{ i \hat H \tau} 
   \hat \sigma_x \mathrm{e}^{-i \hat H \tau} |{-y}\rangle \nonumber \\ 
 &=\langle {-y}| \Bigg(    \cos \Big( \frac{\tau}{2}\sqrt{\Delta B_{z}^2 + J^2} \Big) \mathds{1} \nonumber \\ 
                     &+i \sin \Big( \frac{\tau}{2}\sqrt{\Delta B_{z}^2 + J^2} \Big) \frac{\Delta B_{z}}{\sqrt{\Delta B_{z}^2 + J^2}} \hat \sigma_x \nonumber \\ 
                     &+i \sin \Big( \frac{\tau}{2}\sqrt{\Delta B_{z}^2 + J^2} \Big)
                         \frac{J}{\sqrt{\Delta B_{z}^2 + J^2}} \hat \sigma_z \Bigg) \nonumber \\ 
                     &\times \hat \sigma_x
               \Bigg(    \cos \Big( \frac{\tau}{2}\sqrt{\Delta B_{z}^2 + J^2} \Big) \mathds{1} \nonumber \\ 
                     &-i \sin \Big( \frac{\tau}{2}\sqrt{\Delta B_{z}^2 + J^2} \Big) \frac{\Delta B_{z}}{\sqrt{\Delta B_{z}^2 + J^2}} \hat \sigma_x \nonumber \\ 
                     &-i \sin \Big( \frac{\tau}{2}\sqrt{\Delta B_{z}^2 + J^2} \Big)
                         \frac{J}{\sqrt{\Delta B_{z}^2 + J^2}} \hat \sigma_z \Bigg) |{-y}\rangle \nonumber \\ 
&=\frac{J}{\sqrt{\Delta B_{z}^2 + J^2}} \sin \Big( \sqrt{\Delta B_{z}^2 + J^2} \tau \Big).
\end{align}

\begin{align}
 \langle \hat \sigma_y^{\mathrm{FID}}(\tau) \rangle &= -\cos\Big[ \sqrt{\Delta B_{z}^2 +J^2} \tau \Big] \\
 & \approx - \cos\Big[ \left( J + \frac{\Delta B_{z}^2}{2J} \right) \tau \Big],
\end{align}
the approximation is good for $\tau \ll \frac{8 J^3}{\Delta B_{z}^4} $.

\begin{equation}
 \langle \hat \sigma_z^{\mathrm{FID}}(\tau) \rangle = - \frac{\Delta B_{z}}{\sqrt{\Delta B_{z}^2 +J^2}} 
                                               \sin \Big( \sqrt{\Delta B_{z}^2 +J^2} \tau \Big).
\end{equation}


In the ideal case, SE signals 
(i.e.~average $\langle \hat \sigma_i^{\mathrm{SE}}(\tau) \rangle$ components  of $S$-$T_{0}$ qubit)
evolve as follows.

\begin{align}
  \langle \hat \sigma_x^{\mathrm{SE}}(\tau) \rangle
  &= \langle {-y}|       \mathrm{e}^{ i \hat H \frac{\tau}{2}}
     ( i \hat \sigma_x)\mathrm{e}^{ i \hat H \frac{\tau}{2}}
         \hat \sigma_x
  \nonumber \\
  & \times
  \mathrm{e}^{-i \hat H \frac{\tau}{2}}
     (-i \hat \sigma_x)\mathrm{e}^{-i \hat H \frac{\tau}{2}}
     |{-y}\rangle
      \nonumber \\
  &=\frac{8 \Delta B_{z}^2 J}{(\Delta B_{z}^2 +J^2)^{3/2}}
   \cos  \Big( \frac{1}{4} \sqrt{\Delta B_{z}^2 +J^2} \tau \Big)
    \nonumber \\
 &\times \sin^2\Big( \frac{1}{4} \sqrt{\Delta B_{z}^2 +J^2} \tau \Big).
\end{align}

\begin{align}
 \langle \hat \sigma_y^{\mathrm{SE}}(\tau) \rangle
 &= \frac{1}{\Delta B_{z}^2 + J^2}
 \nonumber \\
 &\times
 \Big( J^2 
 + \Delta B_{z}^2 \cos \Big( \sqrt{\Delta B_{z}^2 +J^2} \tau \Big) \Big).
\end{align}

\begin{align}
 \langle \hat \sigma_z^{\mathrm{SE}}(\tau) \rangle
 &= \frac{\Delta B_{z}}{(\Delta B_{z}^2 + J^2)^{3/2}}
  \nonumber \\
 &\times
 \left[ 2J^2 \sin^2 \left( \frac{1}{2} \sqrt{\Delta B_{z}^2 +J^2} \tau \right) \right.
   \nonumber \\
 &
 + \left. \Delta B_{z}^2 \sin \left( \sqrt{\Delta B_{z}^2 +J^2} \tau \right)  \right].
\end{align}

Assuming that $\bar J \gg \overline{\Delta B_{z}}, \, \sigma_{J}$,
after averaging over quasistatic fluctuations of the parameter $\Delta B_{z}$ or $J$
one obtains following approximate expressions for $y$ qubit component during SE.
\begin{align}
 \langle \langle \hat \sigma_y^{\mathrm{SE}}(\tau) \rangle \rangle_{\Delta B_{z}}
 &\approx
 \frac{J^2}{J^2+\overline{\Delta B_{z}}^2} \nonumber \\
 &+ \frac{1}{J^2+\overline{\Delta B_{z}}^2}
 \exp\left( - \frac{\overline{\Delta B_{z}}^2 \sigma_{\Delta B_{z}}^2 \tau^2}{2(J^2+(\sigma_{\Delta B_{z}}^2 \tau)^2)} \right) \nonumber \\
 &\times \Bigg[ \frac{J^{3/2}(J(\overline{\Delta B_{z}}^2 + \sigma_{\Delta B_{z}}^2) + i \sigma_{\Delta B_{z}}^4 \tau)}
                     {2(J + i \sigma_{\Delta B_{z}}^2 \tau)^{5/2}} \Bigg. \nonumber \\
 &\times \exp\left( -i J \tau \frac{\overline{\Delta B_{z}}^2 +2J^2 + 2(\sigma_{\Delta B_{z}}^2 \tau)^2}{2(J^2+(\sigma_{\Delta B_{z}}^2 \tau)^2)} \right)
 \nonumber \\
 &\Bigg. + \, \mathrm{c.c.} \Bigg].
\end{align}

\begin{align}
 \langle \langle \hat \sigma_y^{\mathrm{SE}} (\tau) \rangle \rangle_{J}
 &\approx \frac{\bar J^2 + \sigma_J^2}{\bar J^2 + \Delta B_{z}^2}
 \nonumber \\
 &
 +\frac{\Delta B_{z}^2}{\bar J^2 + \Delta B_{z}^2} \exp\left(- \frac{\sigma_J^2 \tau^2}{2}\right)
 \cos \left( \bar J \tau \right).
\end{align}

\section{$S$-$T_{0}$ Qubit Attenuation Factors Derived for Dynamically Fluctuating Exchange Splitting}  \label{app:chi}
We present the calculation of
the attenuation factors $\chi_{\mathrm{FID}}(\tau)$ and $\chi_{\mathrm{SE}}(\tau)$ 
that account for the effect of dynamically fluctuating exchange splitting $J$.
\begin{widetext}
The attenuation factor $\chi_{\mathrm{FID}}(\tau)$ is calculated as follows.
\begin{flalign}
 \chi_{\mathrm{FID}}(\tau)
 &= \int_0^{\tau} \mathrm{d} t_1 \int_0^{\tau} \mathrm{d} t_2 \langle \delta J (t_1) \delta J (t_2) \rangle f_{\mathrm{FID}}(t_1) f_{\mathrm{FID}}(t_2)
 \nonumber \\
 &= \int_{-\infty}^{\infty} \mathrm{d} t_1 \int_{-\infty}^{\infty} \mathrm{d} t_2
    \int_{-\infty}^{\infty} \frac{\mathrm{d} \omega}{2\pi} \mathrm{e}^{-i \omega (t_1 - t_2)} S(\omega)
 \nonumber \\
 &\times
    \int_{-\infty}^{\infty} \frac{\mathrm{d}\omega_1}{2\pi} \tilde f_{\mathrm{FID}}(\omega_1) \mathrm{e}^{-i \omega_1 t_1}
    \int_{-\infty}^{\infty} \frac{\mathrm{d}\omega_2}{2\pi} \tilde f_{\mathrm{FID}}(\omega_2) \mathrm{e}^{-i \omega_2 t_2}
 \nonumber \\
 &= \int_{-\infty}^{\infty} \mathrm{d} \omega_1
    \int_{-\infty}^{\infty} \mathrm{d} \omega_2
    \int_{-\infty}^{\infty} \frac{\mathrm{d} \omega}{2\pi} \delta(\omega + \omega_1) \delta(\omega - \omega_2)
 \nonumber \\
 &\times
    \tilde f_{\mathrm{FID}}(\omega_1)  \tilde f_{\mathrm{FID}}(\omega_2) S(\omega)
 \nonumber \\
 &= \int_{-\infty}^{\infty} \frac{\mathrm{d}\omega}{2\pi}  \tilde f_{\mathrm{FID}}(-\omega)  \tilde f_{\mathrm{FID}}(\omega) S(\omega)
  = \int_{-\infty}^{\infty} \frac{\mathrm{d}\omega}{2\pi} \left|  \tilde f_{\mathrm{FID}}(\omega) \right|^2 S(\omega)
 \nonumber \\
 &= \int_{-\infty}^{\infty} \frac{\mathrm{d}\omega}{2\pi} \frac{2 F_{\mathrm{FID}}(\omega \tau)}{\omega^2} S(\omega)
  = 2 \int_{0}^{\infty} \frac{\mathrm{d}\omega}{\pi} \frac{F_{\mathrm{FID}}(\omega \tau)}{\omega^2} S(\omega).
\end{flalign}

Using the explicit analytical expressions for the filter function
$F_{\mathrm{FID}}(\omega \tau)
= 2 \sin^2 \frac{\omega \tau}{2}$
and the spectral density of the noise
$S(\omega)
 = \frac{A}{\omega^{\beta}}$, 
where $A$ is a constant corresponding to the power of the noise,
we obtain the following expression for the attenuation factor
\begin{equation}
\chi_{\mathrm{FID}}(\tau)
 = \frac{4A}{\pi} \int_0^{\infty} \mathrm{d} \omega
 \sin^2 \frac{\omega \tau}{2} \frac{1}{\omega^{2+\beta}},
\end{equation}
which for $\beta = 0.7$ can be evaluated giving finally:
\begin{equation}
\chi_{\mathrm{FID}}(\tau)
 =\frac{2A}{\pi}
 \cos\left( \frac{3 \pi}{20} \right)
 \Gamma \left( -1.7 \right)
 \tau^{1.7}
 \approx 2.24 \, \frac{2A}{\pi} \tau^{1.7},
\end{equation}
as the definite integral can be calculated analytically
\begin{equation}
 \int_0^{\infty} \mathrm{d} \omega 
 \sin^2 \frac{\omega \tau}{2} \frac{1}{\omega^{2+0.7}}
 = \frac{1}{2}
    \cos\left( \frac{3 \pi}{20} \right)
    \Gamma \left( -1.7 \right)
    \tau^{1.7}.
\end{equation}

Similarly, the attenuation factor $\chi_{\mathrm{SE}}(\tau)$ reads as follows.
\begin{flalign}
 \chi_{\mathrm{SE}}(\tau)
 &= \int_0^{\tau} \mathrm{d} t_1 \int_0^{\tau} \mathrm{d} t_2 \langle \delta J (t_1) \delta J (t_2) \rangle f_{\mathrm{SE}}(t_1) f_{\mathrm{SE}}(t_2)
 \nonumber \\
 &= 2\int_{0}^{\infty} \frac{\mathrm{d}\omega}{\pi} \frac{F_{\mathrm{SE}}(\omega \tau)}{\omega^2} S(\omega).
\end{flalign}

Using the explicit analytical expressions for the filter function
$F_{\mathrm{SE}}(\omega \tau)
= 8 \sin^4 \frac{\omega \tau}{4}$
and the spectral density of the noise
$S(\omega)
= \frac{A}{\omega^{\beta}}$,
we obtain the following expression for the attenuation factor
\begin{equation}
 \chi_{\mathrm{SE}}(\tau)
 = \frac{16 A}{\pi} \int_0^{\infty} \mathrm{d} \omega
 \sin^4 \frac{\omega \tau}{4} \frac{1}{\omega^{2+\beta}},
\end{equation}
which for $\beta = 0.7$ can be evaluated giving finally:
\begin{equation}
\chi_{\mathrm{SE}}(\tau)
 = \frac{2 A}{\pi}
    \left(2^{0.3} -1\right)
    \cos\left(\frac{3 \pi}{20}\right)
    \Gamma\left(-1.7\right) 
    \tau^{1.7}
  \approx 0.52 \, \frac{2 A}{\pi} \tau^{1.7},
\end{equation}
as the definite integral can be calculated analytically
\begin{equation}
 \int_0^{\infty} \mathrm{d} \omega \sin^4 \frac{\omega \tau}{4} \frac{1}{\omega^{2+0.7}}
 = \frac{1}{8} 
    \left( 2^{0.3} - 1 \right)
    \cos \left( \frac{3 \pi}{20} \right)
    \Gamma \left( -1.7 \right) 
    \tau^{1.7}.
\end{equation}

The attenuation factor $\chi_{\mathrm{SE}}(\tau)$ for noise with $\beta = 2.0$ is $\chi_{\mathrm{SE}}(\tau) = \frac{A}{12} \tau^3$
as the following definite integral can be calculated analytically
$
\int_0^{\infty} \mathrm{d} \omega \sin^4 \frac{\omega \tau}{4} \frac{1}{\omega^{2+2.0}}
= \frac{\pi}{192} \tau^3
$.


\end{widetext}
%

\end{document}